\documentclass[aps, pra, reprint,preprintnumbers,amsmath,amssymb,superscriptaddress,nofootinbib,longbibliography]{revtex4-2}

\usepackage[utf8]{inputenc}
\usepackage[english]{babel}

\usepackage{graphicx}

\usepackage{leftidx}

\usepackage{diffcoeff}

\usepackage{amsfonts,amssymb,amsmath}
\usepackage{mathtools}

\usepackage{bm}
\usepackage{times}
\usepackage{bbm}
\usepackage{siunitx}
\usepackage{mathrsfs}

\usepackage{array}
\usepackage{tabularx}
\usepackage{multirow}

\usepackage{hyperref}
\hypersetup{colorlinks=true,linktoc=all,linkcolor=blue,breaklinks=true,citecolor=blue,urlcolor=blue}

\usepackage[normalem]{ulem}
\usepackage{xspace}
\usepackage[dvipsnames]{xcolor}

\addto\captionsenglish{}

\newcommand{\e}{\text{e}}
\newcommand{\Ree}{\mathrm{Re}}
\newcommand{\Imm}{\mathrm{Im}}
\renewcommand*{\L}{\text{L}}
\newcommand*{\R}{\text{R}}
\newcommand*{\xzpf}{x_\text{zpf}}
\newcommand*{\las}{\text{las}}
\newcommand*{\alas}{\alpha_\las}
\newcommand*{\kL}{\kappa_\L}
\newcommand*{\kR}{\kappa_\R}
\newcommand*{\kd}{\kappa_d}
\newcommand*{\ka}{\kappa_a}
\newcommand*{\oma}{\omega_a}
\newcommand*{\omd}{\omega_d}
\newcommand*{\DaL}{\Delta_{a_\L}}
\newcommand*{\DaR}{\Delta_{a_\R}}
\newcommand*{\Dd}{\Delta_d}
\newcommand*{\dD}{\delta\Delta}

\newcommand{\ba}{\bar{a}}
\newcommand{\baL}{\ba_\L}
\newcommand{\baR}{\ba_\R}
\newcommand{\bd}{\bar{d}}
\newcommand{\bq}{\bar{q}}
\newcommand{\p}{\hat{p}}
\newcommand{\q}{\hat{q}}
\renewcommand{\a}{\hat{a}}
\renewcommand{\d}{\hat{d}}
\newcommand{\da}{\delta\a}
\newcommand{\dd}{\delta\d}
\newcommand{\dq}{\delta\q}
\newcommand{\deltp}{\delta\p}
\newcommand{\dX}{\delta\hat{X}}
\newcommand{\dP}{\delta\hat{P}}
\newcommand*{\bin}[1][\ell]{\hat{b}_{\mathrm{in},#1}}
\newcommand*{\bout}[1][\ell]{\hat{b}_{\mathrm{out},#1}}
\newcommand*{\ain}[1][\ell]{\hat{a}_{\mathrm{in},#1}}
\newcommand*{\ainL}{\ain[\L]}
\newcommand*{\ainR}{\ain[\R]}
\newcommand*{\din}{\hat{d}_\text{in}}
\newcommand*{\gam}{\gamma_\mathrm{m}}
\newcommand*{\Om}{\Omega_\mathrm{m}}
\newcommand{\nth}{\bar{n}_\text{m}}
\newcommand{\nfin}{\bar{n}_\text{fin}}
\newcommand*{\aL}{\a_\L}
\newcommand*{\aR}{\a_\R}
\newcommand*{\daL}{\da_\L}
\newcommand*{\daR}{\da_\R}
\newcommand*{\gaL}{g_{a_\L}}
\newcommand*{\gaR}{g_{a_\R}}

\newcommand*{\N}{\text{N}}
\newcommand*{\F}{\text{F}}
\newcommand*{\D}{\text{D}}
\newcommand*{\thN}{\theta_\N}
\newcommand*{\thD}{\theta_\D}
\newcommand*{\gF}{\gamma_\F}
\newcommand*{\omF}{\omega_\F}
\newcommand*{\zN}{\zeta_\N}
\newcommand*{\zD}{\zeta_\D}

\newcommand*{\bM}{{\bm M}}
\newcommand*{\TM}{{\text{TM}}}
\newcommand*{\RM}{{\text{RM}}}
\newcommand*{\TMRM}{{\TM/\RM}}
\newcommand*{\tm}{{M}}
\newcommand*{\FSR}{\Gamma_\text{FSR}}
\newcommand{\figpanel}[2]{\hyperref[#1]{\ref*{#1}(#2)}}

\DeclarePairedDelimiter{\abs}{\lvert}{\rvert}
\DeclarePairedDelimiter{\mean}{\langle}{\rangle}

\newcommand{\subfigref}[2]{\ref{#1}\hyperref[#1]{(#2)}}

\begin{document}
    
\title{Optomechanical systems with a Fano membrane in the middle}

\date{\today}

\author{Lei Du}
\email{lei.du@chalmers.se}
\affiliation{Department of Microtechnology and Nanoscience (MC2), Chalmers University of Technology, 412 96 Gothenburg, Sweden}
\author{Aymeric Frerejean}
\affiliation{Department of Microtechnology and Nanoscience (MC2), Chalmers University of Technology, 412 96 Gothenburg, Sweden}
\author{Witlef Wieczorek}
\affiliation{Department of Microtechnology and Nanoscience (MC2), Chalmers University of Technology, 412 96 Gothenburg, Sweden}
\author{Janine Splettstoesser}
\affiliation{Department of Microtechnology and Nanoscience (MC2), Chalmers University of Technology, 412 96 Gothenburg, Sweden}
\author{Juliette Monsel}
\affiliation{Department of Microtechnology and Nanoscience (MC2), Chalmers University of Technology, 412 96 Gothenburg, Sweden}

\begin{abstract}
Conventional membrane-in-the-middle (MIM) optomechanical systems offer limited control over the optical linewidth, which can constrain their performance when operating in the unresolved-sideband regime. We investigate cavity optomechanics with a photonic-crystal Fano membrane placed at the center of a Fabry--Pérot (FP) cavity. In contrast to a conventional dielectric membrane, the photonic-crystal membrane supports a localized optical resonance, which hybridizes with the cavity field, enabling spectral engineering of the relevant optical modes. Besides the usual dispersive optomechanical coupling associated with cavity-length changes, the membrane motion also modifies the Fano-mode resonance and its hybridization with the cavity field. We consider two limits set by the membrane reflectivity: a transparent-membrane regime with a single FP-like mode, and a reflective-membrane regime with two coupled subcavity modes. In the latter case, only the symmetric cavity mode hybridizes with the Fano mode, while the antisymmetric mode remains decoupled. Using quantum Langevin equations together with a transfer-matrix description of the optical scattering problem, we show that the Fano-induced hybridization can generate narrow optical normal modes that remain efficiently accessible to the external drive for experimentally realistic parameters. These modes can provide effective sideband resolution and enable ground-state cooling of the membrane motion even when the bare cavity is in the unresolved-sideband regime. Our results establish Fano MIM systems as a promising platform for spectral and optomechanical engineering.
\end{abstract}

\maketitle
    
\section{Introduction}

Cavity optomechanics provides a promising and versatile platform for studying and controlling the interaction between electromagnetic fields and mechanical motion~\cite{Aspelmeyer2014Dec,Kippenberg2007Review,OMQT,OConnell2010}. Radiation-pressure interactions enable precision sensing~\cite{OMmeasure1,OMmeasure2,OMmeasure3,Krause2012sensing}, coherent state transfer and frequency conversion~\cite{OMtransfer,Fconversion1,Fconversion2,Fconversion3,Fconversion4}, as well as the preparation of nonclassical mechanical states~\cite{Bose1997PRA,Marshall2003PRL,RablPB2011,SVCR2,Garziano2015PRA,Hauer2023prl}. A central objective is to cool a mechanical resonator close to its quantum ground state~\cite{Genes2008Mar,MBF4,Teufel2011Jul,SidebandCooling2011TK,Delic2020Feb}, which is generally achieved by driving an optical resonance on its red mechanical sideband. Efficient sideband cooling requires the resolved-sideband regime~\cite{Aspelmeyer2014Dec,Kippenberg2007Review,OMQT}, in which the relevant optical linewidth is smaller than the mechanical frequency, allowing anti-Stokes scattering to be spectrally selected over Stokes scattering.

Membrane-in-the-middle (MIM) systems are among the most widely studied implementations of cavity optomechanics~\cite{Paternostro2007Dec, Thompson2008Mar,Jayich2008Sep,Karuza2012MIM, Aspelmeyer2014Dec, Enzian2024May}. In their conventional form, a thin dielectric membrane is placed inside a Fabry--Pérot (FP) cavity, where its displacement shifts the cavity resonance frequencies and thereby generates dispersive optomechanical coupling. This architecture spatially separates the optical and mechanical functionalities, allowing the optical and mechanical properties to be engineered with substantial independence~\cite{Thompson2008Mar,Jayich2008Sep,Enzian2024May,Karuza2012MIM}. It has further enabled tunable nonlinear and quadratic optomechanical interactions~\cite{Thompson2008Mar,Sankey2010Sep,Nunnenkamp2010MIM,Burgwal2020Nov, Ya-FengJiao2025Oct}, including regimes relevant for quantum nondemolition measurements of mechanical energy~\cite{Sankey2010Sep}. Extensions to multiple membranes can additionally enhance the coupling to collective mechanical modes through optical interference~\cite{Xuereb2012PRL}. Depending on the membrane reflectivity and position, the optical field can either extend across the full cavity \cite{Ya-FengJiao2025Oct} or be described in terms of coupled subcavity modes~\cite{Paternostro2007Dec, Jayich2008Sep, Enzian2024May}. However, conventional dielectric membranes mainly act as frequency-independent scatterers and therefore offer limited control over the spectral properties of the relevant optical modes. In particular, they do not provide a narrow internal optical resonance that can strongly reduce the effective optical linewidth. As a consequence, the optical response is typically governed by the bare cavity linewidth, which can preclude resolved-sideband operation in lossy cavities and thereby hinder ground-state cooling and other quantum-control protocols.

A promising route to overcome this limitation is to use a photonic-crystal membrane with a strongly \emph{frequency-dependent} optical response, which can support localized optical resonances and Fano-type spectral features in reflection and transmission~\cite{Fan2003Mar,Limonov2017Sep,Suh2005Aug}. When incorporated into a cavity optomechanical system, the localized membrane resonance, hereafter referred to as the ``Fano mode'', can hybridize with a FP-like cavity mode. Under suitable conditions, this hybridization produces optical normal modes with linewidths substantially narrower than those of the bare optical modes, thereby enabling effective sideband resolution even when the bare cavity lies deep in the unresolved-sideband regime~\cite{Cernotik2019Jun,WWoe2023,Monsel2024Apr,DL2025Fano,Du2026QST}. As a result, such Fano-membrane architectures provide a promising route towards ground-state cooling in microcavity optomechanics~\cite{Monsel2024Apr,DL2025Fano} and access to nonlinear quantum optomechanical regimes~\cite{Fitzgerald2021Feb,Du2026QST}.

\begin{figure}[tb]
    \includegraphics[width=\linewidth]{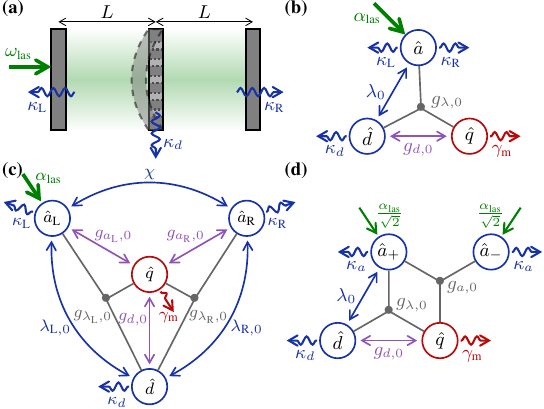}\vspace{-0.2cm}
    \caption{\label{fig:Model}
        (a) Schematics of the membrane-in-the-middle setup. (b) Diagram of the system in the transparent-membrane limit where only one cavity mode ($\a$) is relevant, coupling to the optical membrane mode ($\d$) and to the mechanical mode ($\q$), see Sec.~\ref{sec:modelA}. (c) Diagram of the system in the reflective-membrane limit, with  a left cavity mode ($\aL$), a right cavity mode ($\aR$) and the optical membrane mode ($\d$), coupled to each other and to the mechanical mode ($\q$). (d) Diagram of the system in the reflective-membrane limit in the symmetric case where the left and right cavity modes are identical, in the $\{\a_-,\,\a_+,\,\d\}$ basis, see Sec.~\ref{sec:modelB}.
    }\vspace{-0.2cm}
\end{figure}

Previous theoretical and experimental studies of Fano-membrane optomechanics have, to our knowledge, mainly focused on geometries in which the frequency-dependent membrane serves as one end mirror of a microcavity~\cite{Cernotik2019Jun,WWoe2023,Monsel2024Apr,DL2025Fano,Du2026QST}. In this configuration, the Fano mode hybridizes with a single cavity mode through both coherent coupling and collective dissipation mediated by a shared photonic reservoir. Placing a Fano membrane inside a cavity instead poses a qualitatively distinct problem. The membrane then acts simultaneously as a movable scatterer, a frequency-dependent optical element, and, at sufficiently high reflectivity, as a shared boundary separating two optical subcavities. Consequently, the relevant mode structure, the symmetry of the optical hybridization, and the resulting optomechanical interactions all strongly depend on the membrane reflectivity. It is therefore not evident whether, or under which conditions, the linewidth-narrowing mechanism known from end-mirror Fano systems can be retained and exploited in a MIM geometry.
Furthermore, recent works~\cite{DeJong2022Feb, Enzian2024May, Yao2025Jan} have put forward photonic-crystal-patterned membranes for MIM optomechanics due to their enhanced reflectivity. However, they do not account for the frequency-dependent optical response and Fano-mode hybridization that such patterning can introduce in the modeling, although Ref.~\cite{Yao2025Jan} notes a discrepancy between the model and the measured data and suggests other sources of optomechanical couplings could explain it.

In this paper, we therefore investigate cavity optomechanics with a photonic-crystal Fano membrane positioned at the center of a FP cavity. We focus on two practical regimes determined by the membrane reflectivity. In the transparent-membrane (TM) limit, the intracavity field remains delocalized over the total cavity length and is described by a single FP-like mode coupled to the localized Fano mode. In the reflective-membrane (RM) limit, the membrane partitions the cavity into left and right subcavities, giving rise to two FP-like modes that interact with the Fano mode and, for finite membrane transmissivity, with each other through photon tunneling. The membrane displacement affects the optical subsystem through both the shifts of the optical resonances and the displacement dependence of the coupling between the FP-like and Fano modes.

We study these regimes using two complementary approaches. The \emph{quantum Langevin-equation framework} captures the driven-dissipative optomechanical dynamics and allows us to determine whether the resulting optical normal modes can be exploited for ground-state cooling once mechanical motion and noise are included~\cite{Aspelmeyer2014Dec,Kippenberg2007Review, Monsel2024Apr}. The \emph{transfer-matrix method}~\cite{Klein1986, Deutsch1995Aug, Xuereb2009May, Xuereb2012NJP, Cernotik2019Jun}, on the other hand, treats the underlying optical scattering problem: it identifies the relevant optical resonances and determines how their frequencies, linewidths, and effective coupling parameters depend on the membrane reflectivity, thereby providing a natural connection between the TM and RM limits~\cite{Fan2003Mar,Jayich2008Sep,Enzian2024May}. Together, these approaches reveal how Fano-induced optical hybridization can be used to engineer narrow, optically accessible normal modes and to enhance control of the mechanical motion in regimes where bare sideband resolution is challenging. Finally, we show that ground-state cooling is achievable with experimentally realistic parameters, paving the way towards the quantum regime and sensing applications.

The remainder of the paper is organized as follows. First, we introduce the MIM system and present the quantum coupled-mode models and Langevin-equation framework describing it in the TM (Sec.~\ref{sec:modelA}) and RM (Sec.~\ref{sec:modelB}) limits. Then, in Sec.~\ref{sec:cooling}, we show that the engineering of the Fano membrane allows for ground-state cooling. Finally, in Sec.~\ref{sec:exp}, we establish the experimental feasibility of our scheme, connecting the parameters of the coupled-mode models to the device's geometry and the mirrors' and membrane's reflectivity via a transfer-matrix approach, which is presented in detail in Appendix \ref{app:transfer matrix}.

\section{Membrane-in-the-middle systems}\label{SecModel}

As depicted in Fig.~\figpanel{fig:Model}{a}, we consider a FP optical cavity of total length $2L$, with a photonic-crystal membrane positioned at its midpoint. In contrast to conventional MIM optomechanical systems based on dielectric membranes~\cite{Thompson2008Mar,Jayich2008Sep,Karuza2012MIM,Enzian2024May,Sankey2010Sep,Nunnenkamp2010MIM,Burgwal2020Nov}, the photonic-crystal membrane considered here supports two relevant degrees of freedom: a mechanical vibrational mode with frequency $\Om$ and damping rate $\gam$, and a localized optical mode, denoted $\d$, with resonance frequency $\omd$ and intrinsic loss rate $\kd$. Hereafter in this paper, we refer to this localized optical mode as the ``Fano mode'' and therefore the membrane as the ``Fano membrane''. The cavity is driven by a laser of frequency $\omega_\las$ through the left mirror.

The optical description of the system depends on the reflectivity of the Fano membrane. Specifically, the relevant FP-like optical field can be modeled either as a single cavity mode $\a$ (in the TM case where the membrane is nearly transparent), or as two cavity modes $\aL$ and $\aR$ for the left and right subcavities (in the RM case where the membrane is highly reflective). In both limits, the FP-like mode or modes couple coherently to the Fano mode with strength $\lambda$, due to the spatial overlap of their fields \cite{Cernotik2019Jun}. In the RM case, photons can also tunnel directly between the two subcavity modes $\aL$ and $\aR$ at a rate $\chi$. 
The mechanical motion of the membrane affects the optical degrees of freedom through different mechanisms in the two reflectivity limits. In the TM case, the optomechanical coupling between the mechanical displacement $\q$ and the FP-like cavity mode $\a$ is strongly suppressed, because the membrane is nearly transparent and therefore exhibits a negligible radiation-pressure effect. In the RM case, by contrast, the membrane acts as a shared movable boundary for the two subcavities. Its displacement increases the length of one subcavity while decreasing that of the other, giving rise to dispersive optomechanical couplings $g_{a_\L,0}$ and $g_{a_\R,0}$ with opposite signs \cite{Aspelmeyer2014Dec, Burgwal2020Nov}.

Moreover, the mechanical vibration also modifies the optical properties of the localized Fano mode. In particular, the out-of-plane motion of the Fano membrane can induce in-plane strain, which shifts the Fano-mode resonance frequency and changes its optical linewidth \cite{WWoe2023, Monsel2024Apr}. The frequency shift gives rise to a dispersive optomechanical coupling \(g_{d,0}\) between the mechanical mode and the Fano mode. As discussed below, the displacement dependence of the Fano optical response can also induce a correction to the coherent coupling between the cavity and Fano modes, making \(\lambda\) effectively dependent on the mechanical displacement.

In the following sections, we discuss the TM (Sec.~\ref{sec:modelA}) and RM  (Sec.~\ref{sec:modelB}) limits in more detail and formulate the corresponding Hamiltonians and equations of motion.

\subsection{Transparent-membrane case}\label{sec:modelA}

\subsubsection{Hamiltonian and dynamics}

As discussed above and illustrated in Fig.~\figpanel{fig:Model}{b}, in the TM regime, the intracavity field can be well described by a single FP-like mode $\a$ with resonance frequency $\oma$.
The localized membrane mode (i.e., the Fano mode) $\d$ couples coherently to the cavity mode $\a$ with rate $\lambda$ and dispersively to the mechanical motion with single-photon coupling strength $g_{d,0}$.
In the frame rotating at the laser frequency $\omega_\las$, the Hamiltonian is\vspace{-0.05cm}
\begin{align}
\!\!\frac{\hat{H}_\TM}{\hbar} &=
\Delta_{a,0} \a^\dagger \a + \Delta_{d,0} \d^\dagger \d
+\frac{\Om}{2}\left(\p^2+\q^2\right) \nonumber \\
&\quad\, +\lambda_0\left(\a^\dagger \d + \d^\dagger \a\right) + i\sqrt{2\kappa_{\rm L}}\alas\left(\a^{\dag}-\a\right)
\nonumber\\
&\quad\, -\sqrt{2}g_{d,0}\,\d^\dagger \d\,\q +\sqrt{2}g_{\lambda,0}(\a^\dagger\d + \d^\dagger\a)\q 
,
\label{eq:H_modelA}
\end{align}
where $\q$ and $\p$ are the dimensionless position and momentum quadrature operators of the mechanical mode, respectively, satisfying the commutation relation $[\q, \p]=i$. Here, \(\Delta_{a,0}=\oma-\omega_\las\) is the detuning between the FP-like cavity resonance and the laser frequency, while \(\Delta_{d,0}=\omd-\omega_\las\) is the corresponding detuning for the Fano mode. $\alas$ is the coherent drive amplitude injected from the left. We denote by $\kappa_{\rm L}$ the decay rate of the cavity mode through the left mirror. As mentioned above, the mechanical displacement also modifies the optical properties of the Fano mode and thereby modulates the optical coupling $\lambda$. We account for this effect by assuming $\lambda(\q)\simeq\lambda_0+\sqrt{2}g_{\lambda,0}\q$, where $\lambda_0$ is the constant baseline and $g_{\lambda,0}$ quantifies the strength of this optical-coupling modulation.  

Including damping and input noise  \cite{Gardiner1985Jun, Genes2008Mar, Monsel2024Apr}, the dynamics of the system is described by the Heisenberg--Langevin equations
\begin{subequations}\allowdisplaybreaks
\label{eq:QLE_modelA}
\begin{align}
\dot{\a} &= -(i\Delta_{a,0}+\kappa_a)\a 
-i\lambda_0 \d -i\sqrt{2}g_{\lambda, 0}\q\d 
\nonumber\\
&\quad\, + \sqrt{2\kR}\ainR +\sqrt{2\kL}\left(\alas+\ainL\right),\\
\dot{\d} &= -(i\Delta_{d,0}+\kd)\d -i\lambda_0 \a+ i\sqrt{2}g_{d,0}\q\,\d
  \nonumber\\
&\quad\, -i\sqrt{2}g_{\lambda, 0}\q\a + \sqrt{2\kd}\,\din,\\
\dot{\q} &= \Om\p,\\
\dot{\p} &= -\Om\q-\gam\p
 -\sqrt{2}g_{\lambda,0}(\a^\dagger\d + \d^\dagger\a) \nonumber\\
&\quad\, +\sqrt{2}g_{d,0}\d^\dagger \d
+\sqrt{\gam}\,\hat\xi.
\end{align}
\end{subequations}
Here, $\kappa_{\rm R}$ denotes
the decay rate of the cavity mode through the right mirror (similar to $\kappa_{\rm L}$), and $\kappa_a=\kL+\kR$ is the total cavity decay rate.  In the following, we assume identical mirrors, such that $\kL = \kR = \kappa_a/2$. Although the Fano mode is not directly coupled to the left and right photonic reservoirs in the present configuration, we nevertheless include a weak intrinsic loss channel characterized by the decay rate $\kd$, corresponding to, e.g., absorption and scattering losses. The input operators $\ain$, with $\ell=\L,\R$, and $\din$ describe the vacuum noises entering through the two cavity mirrors and the intrinsic loss channel of the Fano mode. Their only nonvanishing correlation functions are $\mean{\ain(t)\ain^\dagger(t')} = \delta(t - t')$ and $\mean{\din(t)\din^\dagger(t')} = \delta(t - t')$. $\hat\xi$ denotes the mechanical thermal noise, with the correlation function\footnote{This is a high temperature approximation, since $\hbar\Om \ll k_\text{B} T$ even at cryogenic temperature for mechanical frequencies in the \si{\mega\hertz} range, which is known for giving a spurious term in the phase noise spectrum \cite{Giovannetti2001Jan}, but it does not impact the steady-state phonon number in the mechanical fluctuations that we are interested in. See, for instance, the discussion in the Appendix of Ref.~\cite{Juliette2021pra}.} $\mean{\hat{\xi}(t)\hat{\xi}(t')} \simeq (2\nth + 1)\delta(t - t')$ where $\nth = [\exp(\hbar\Om/k_\text{B} T) - 1]^{-1}$ is the thermal phonon number at temperature $T$.

\subsubsection{Linearized dynamics}\label{sec:lin:modelA}

In this work, we focus on the weak-coupling regime, where the optomechanical interaction can be treated perturbatively, and consider a large laser power $P_\las = \hbar\omega_\las \abs{\alas}^2$. We therefore linearize the dynamics~\cite{Genes2008Mar, Monsel2024Apr} by writing $\a=\ba+\da$, $\d=\bd+\dd$, $\q=\bq+\dq$, and $\p=\bar p+\deltp$, where $\ba=\mean{\a}$, $\bd=\mean{\d}$, $\bq=\mean{\q}$, and $\bar{p} = \mean{\p}$.
Solving the classical steady-state equations yields
\begin{subequations}\allowdisplaybreaks
\label{eq:ss_modelA}
\begin{align}
\ba &=
\frac{\sqrt{\kappa_a}\,(i\Dd+\kd)}
{(i\Delta_{a,0}+\kappa_a)(i\Dd+\kd)+\lambda^2}\,\alas,\\
\bd &=
\frac{-i\sqrt{\kappa_a}\,\lambda}
{(i\Delta_{a,0}+\kappa_a)(i\Dd+\kd)+\lambda^2}\,\alas,\\
\bq &=
\frac{\sqrt{2}}{\Om}
\left[
g_{d,0}|\bd|^2 - g_{\lambda,0}(\ba^*\bd + \bd^*\ba)
\right],
\qquad
\bar p=0,
\end{align}
\end{subequations}
where $\Delta_d=\Delta_{d,0}-\sqrt{2}g_{d,0}\bq$ is the effective detuning of the Fano mode and $\lambda= \lambda_0 + \sqrt{2}g_{\lambda, 0} \bq$ denotes the optical coupling between the Fano mode and the cavity mode evaluated at the mean displacement.
Keeping terms to first order in the fluctuations, the linearized Langevin equations become
\begin{subequations}\allowdisplaybreaks
\label{eq:linQLE_modelA}
\begin{align}
\!\delta\dot{\a}&=-(i\Delta_{a,0}+\kappa_a)\da
-i\lambda\,\dd +i\sqrt{2}g_a\,\dq\nonumber\\
&\quad\,
+\sqrt{\kappa_a}\left(\ainL+\ainR\right), \\
\!\delta\dot{\d}&=-(i\Dd+\kd)\dd
-i\lambda\,\da
+i\sqrt{2}g_d\,\dq
+\sqrt{2\kd}\din,\!\\
\!\delta\dot{\q}&=\Om\deltp,\\
\!\delta\dot{\p}&=-\Om\dq-\gam\deltp + \sqrt{2}\left(g_a\,\da^\dagger+g_a^*\,\da\right)
\nonumber\\
&\quad\,
+\sqrt{2}\left(g_d\,\dd^\dagger+g_d^*\,\dd\right)+\sqrt{\gam}\,\hat\xi,
\end{align}
\end{subequations}
with the laser-enhanced optomechanical coupling strengths $g_a = -g_{\lambda,0}\bd$ and $g_d = g_{d, 0}\bd - g_{\lambda, 0}\ba$. The validity of this linearization for the parameters considered in Sec.~\ref{sec:cooling} is discussed in Appendix~\ref{app:validity}.

To quantify the cooling performance, we evaluate the steady-state fluctuations of the mechanical mode from the linearized quantum Langevin equations. For this purpose, we introduce the optical quadratures \(\dX_c=(\delta\hat c+\delta\hat c^\dagger)/\sqrt{2}\) and \(\dP_c=(\delta\hat c-\delta\hat c^\dagger)/(i\sqrt{2})\) for \(c\in\{a,d\}\). In this quadrature basis, Eq.~\eqref{eq:linQLE_modelA} can be written in a compact matrix form, which allows all steady-state second moments to be obtained by numerically solving the corresponding Lyapunov equation (see Appendix~\ref{app:TM} for details). We then obtain the final phonon occupation with the resulting mechanical quadrature variances, 
\begin{equation}\label{nfin}
    \nfin = \frac{\mean{\dq^2} + \mean{\deltp^2} - 1}{2},
\end{equation}
which serves as our figure of merit for the cooling performance (ground-state cooling occurs if $\bar n_{\rm fin}\ll1$). 
 
\subsubsection{Optical normal modes}\label{sec:modelA:normal modes}

In the parameter regimes considered in this work, the single-photon optomechanical couplings are much weaker than the coherent optical coupling $\lambda_0$ \cite{WWoe2023}. We can therefore first analyze the hybridization of the optical modes in the absence of mechanical motion, and subsequently treat the optomechanical interaction as a perturbation. Neglecting the mechanics, the dynamics of the optical sector is governed by 
\begin{equation}\label{TM_normal_EOM}
\!\!\diff*{\begin{pmatrix}
\a\\ \d
\end{pmatrix}}{t}
=
-i
\begin{pmatrix}
\Delta_{a,0}-i\kappa_a & \lambda_0\\
\lambda_0 & \Delta_{d,0}-i\kd
\end{pmatrix}
\begin{pmatrix}
\a\\ \d
\end{pmatrix}
+\bm D_{\rm TM},
\end{equation}
where $\bm D_{\rm TM}$ contains the drive and noise terms from Eq.~\eqref{eq:QLE_modelA}. Diagonalizing this dynamics gives the complex eigenvalues
\begin{equation}
\Omega^\pm_\TM = \bar{\Delta}_\TM -i\bar{\kappa}\ \pm\ \sqrt{\left(\dD_\TM-i\,\delta\kappa\right)^2+\lambda_0^2},
\label{eq:Omega_pm_modelA}
\end{equation}
where $\bar{\Delta}_\TM=(\Delta_{a,0}+\Delta_{d,0})/2$ and $\bar{\kappa}=(\kappa_a+\kd)/2$ are the average detuning and linewidth, respectively, and
$\dD_\TM=(\Delta_{a,0}-\Delta_{d,0})/2$, $\delta\kappa=(\kappa_a-\kd)/2$ are the corresponding half-differences. 
The corresponding right eigenmodes\footnote{Since the matrix in Eq.~\eqref{TM_normal_EOM} is non-Hermitian for $\ka\neq\kd$, the right eigenmodes are generally non-orthogonal. A complete biorthogonal description~\cite{Biorthogonal2013} would involve both left and right eigenmodes; however, the right eigenmodes are sufficient for characterizing the optical hybridization considered here.}, which we call the optical normal modes, can be written as
\begin{equation}\label{eq:NormalMode_def}
\hat A_\pm
=
\mathcal N_\pm
\left(
\hat a+r_\pm \hat d
\right)
\end{equation}
with
\(r_\pm=
[\Omega^\pm_\TM-(\Delta_{a,0}-i\kappa_a)]/\lambda_0\) and \(\mathcal{N}_\pm
=
1/\sqrt{1+\left|r_\pm\right|^2}\).
The complex coefficients \(r_\pm\) characterize both the relative weight
and the relative phase between the FP-like cavity component and the Fano-mode
component. 
The coherent drive acting on the FP-like cavity mode gives the effective amplitudes
\begin{align}
    \eta_\pm
    &=
   \pm \sqrt{\kappa_a}\alpha_{\rm las}
    \frac{r_\mp}{\mathcal N_\pm\left(r_- - r_+\right)}
\end{align}
in this eigenmode basis.

Equation~\eqref{eq:Omega_pm_modelA} provides a convenient picture of the optical normal modes, in terms of their effective resonance frequencies (detunings) and linewidths,
\begin{equation}
    \Delta^\pm_\TM=\Ree(\Omega^\pm_\TM),\qquad\kappa^\pm_\TM=-\Imm(\Omega^\pm_\TM). \label{DpmTM}
\end{equation}
In the present context, the central question is whether the hybridization can produce optical normal modes that are both sufficiently narrow and accessible to the external drive to enable efficient control of the mechanical motion.

\subsection{Reflective-membrane case}\label{sec:modelB}

\subsubsection{Hamiltonian and dynamics}

In the RM regime, the membrane splits the cavity into two subcavities supporting distinct FP-like modes $\aL$ and $\aR$, with respective bare frequencies $\omega_{a_\L}$, $\omega_{a_\R}$ and loss rates $\kL$, $\kR$, as illustrated in Fig.~\figpanel{fig:Model}{c}.
Mechanical motion of the Fano membrane changes the two subcavity lengths in opposite directions and therefore induces dispersive couplings of opposite signs \cite{Burgwal2020Nov}.
Both subcavity modes couple to the Fano mode $\d$ at rates $\lambda_{\L}$ and $\lambda_{\R}$ (also dependent on $\q$ as discussed below), respectively.
In addition, a small fraction of light may be directly transmitted through the membrane without exciting the membrane mode; this process is described by a direct inter-cavity coupling $\chi$  \cite{Burgwal2020Nov}. 
The Hamiltonian, in the frame rotating at the laser frequency, is given by
\begin{align}
\!\frac{\hat{H}_\RM}{\hbar} =&
\!\!\!\sum_{{c=a_\L,a_\R,d}}\!\!\Delta_{c, 0}\hat{c}^\dagger\hat{c}
+\frac{\Om}{2}\left(\p^2+\q^2\right)+\chi\!\left(\aL^\dagger\aR+\aR^\dagger\aL\right)
\nonumber\\[-0.1cm]
& 
\!+\sum_{{\ell=\L,\R}}\lambda_{\ell,0}\left(\a_{\ell}^\dagger\d+ \d^\dagger\a_{\ell}\right)+ i\sqrt{2\kL}\alas\left(\aL^{\dag}-\aL\right) \nonumber\\[-0.1cm]
&
\!-\sqrt{2}\!\!\!\!\sum_{{c=a_\L,a_\R,d}}\!\!\!g_{c,0}\hat{c}^\dagger\hat{c}\,\q
+\sqrt{2}\!\sum_{{\ell=\L,\R}}\!g_{\lambda_\ell,0}\!\left(\a_{\ell}^\dagger\d + \d^\dagger\a_{\ell}\right)\!\q,\label{eq:H_modelB}\\[-0.65cm]\nonumber
\end{align}
with $\Delta_{c,0}=\omega_{c}-\omega_{\rm las}$ the detuning between the corresponding optical mode and the drive field. Similar to the TM case, we assume $\lambda_{\L}(\q)\simeq\lambda_{\L,0}+\sqrt{2}g_{\lambda_{\L},0}\q$ and $\lambda_{\R}(\q)\simeq\lambda_{\R,0}+\sqrt{2}g_{\lambda_{\R},0}\q$. A representative parameter hierarchy is $\omega_{a_\L}, \omega_{a_\R},\omd \gg \lambda_{\L/\R} \gg \chi$, with the cavity decay rates $\kL, \kR$ on the order of (or moderately larger than) $\Om$, and the intrinsic Fano-mode loss $\kd\lesssim \kL,\kR$.

So far, the model [Eq.~\eqref{eq:H_modelB} and Fig.~\subfigref{fig:Model}{c}] is valid for a general case where the left and right cavities can be different.
From here on, to simplify the analysis and provide a more intuitive picture, we focus on the fully symmetric case where the two end mirrors are identical and the membrane is located  exactly in the middle. As a consequence, we take $\omega_{a_\L} = \omega_{a_\R} \equiv \omega_0$, $\Delta_{a_\L, 0} = \Delta_{a_\R, 0}\equiv \Delta_0$, $\kL = \kR = \ka$, $\lambda_{\L,0} = \lambda_{\R,0} \equiv \lambda_0/\sqrt{2}$, $g_{\lambda_\L,0} = g_{\lambda_\R,0} \equiv g_{\lambda,0}/\sqrt{2}$ and $g_{a_\L,0} = -g_{a_\R,0} \equiv g_{a,0}$. Introducing the symmetric and antisymmetric modes
 $\a_\pm = (\aL \pm \aR)/\sqrt{2}$, with $\Delta_{\pm}=\Delta_{0}\pm\chi$,
 we can rewrite the Hamiltonian from Eq.~\eqref{eq:H_modelB} as
\begin{align}
    \frac{\hat{H}_\RM}{\hbar} &= \sum_{\sigma=\pm}
    \Delta_{\sigma}\a_{\sigma}^\dagger\a_{\sigma} 
    +\Delta_{d, 0}\,\d^\dagger\d +\frac{\Om}{2}\left(\p^2+\q^2\right) \nonumber\\
    &\quad
    +\lambda_0\left(\a_+^\dagger\d+ \d^\dagger \a_+\right)  + i\sqrt{\ka}\alas\sum_{\sigma=\pm}\left(\a_{\sigma}^{\dag}-\a_{\sigma}\right)\nonumber\\
     &\quad 
    -\sqrt{2}g_{d,0}\d^\dagger\d\,\q -\sqrt{2}g_{a,0}\left(\a_+^\dagger\a_- +\a_-^\dagger\a_+\right)\q
   \nonumber\\
   &\quad +\sqrt{2}g_{\lambda,0}\left(\a_+^\dagger\d + \d^\dagger\a_+\right)\q, 
    \label{eq:H_modelB:pm}\\[-0.55cm]\nonumber
\end{align}
which shows a mixed optomechanical interaction among $\a_+$, $\a_-$, and $\q$, as illustrated in Fig.~\figpanel{fig:Model}{d}. Moreover, the antisymmetric normal mode $\a_-$ is decoupled from the Fano mode $\d$.

Including input noise, the equations of motion in this normal-mode basis read (see Appendix~\ref{app:RM} for details)
\begin{subequations}\allowdisplaybreaks
    \label{eq:QLE_modelB:+-}
    \begin{align}
        \dot{\a}_- &= -(i\Delta_- +\ka)\a_-
        +i\sqrt{2}g_{a,0}\q\,\a_+ \\\nonumber
        &\quad\,
        +\sqrt{\ka}\alas+\sqrt{2\ka}\ain[-],\\
        \dot{\a}_+ &= -(i\Delta_+ +\ka)\a_+
        -i\lambda_0\d+i\sqrt{2}g_{a,0}\q\,\a_- \\\nonumber
        &\quad\,-i\sqrt{2}g_{\lambda,0}\q\,\d +\sqrt{\ka}\alas+\sqrt{2\ka}\ain[+],\\
        \dot{\d} &= -(i\Delta_{d,0}+\kd)\d
        -i\lambda_0\a_+ +i\sqrt{2}g_{d,0}\q\,\d\\\nonumber
        &\quad\,-i\sqrt{2}g_{\lambda,0}\q\a_+ +\sqrt{2\kd}\din,\\
        \dot{\q} &= \Om\p,\\
        \dot{\p} &= -\Om\q-\gam\p+\sqrt{2}g_{a,0}\left( \a_+^\dagger\a_- + \a_-^\dagger\a_+\right)\\\nonumber
        &\quad\,
        +\sqrt{2}g_{d,0}\d^\dagger\d-\sqrt{2}g_{\lambda,0}\left(\a_+^\dagger\d+ \d^\dagger\a_+\right)
        +\sqrt{\gam}\,\hat\xi,
    \end{align}
\end{subequations}
where we have defined the input noises $\ain[\pm] = (\ainL \pm \ainR)/\sqrt{2}$.

\subsubsection{Linearized dynamics}\label{sec:lin:modelB}
As in the TM case, we linearize the dynamics around the steady-state mean fields $\baL=\mean{\aL}$, $\baR=\mean{\aR}$, $\bd=\mean{\d}$ and $\bq=\mean{\q}$.
The resulting linearized equations have the same structure as those in the TM case, but exhibit an enlarged optical quadrature space; see Appendix~\ref{app:RM}.
The final phonon occupancy $\nfin$ is again given by Eq.~\eqref{nfin} and computed from the steady-state covariance matrix, obtained by solving a Lyapunov equation; see Appendix~\ref{app:RM}.
In the following, we use this framework to identify parameter regimes in which ground-state cooling of the mechanical mode can be effectively achieved.

\subsubsection{Optical normal modes}\label{sec:modelB:normal modes}

We now analyze the optical normal modes in the RM case. As in the TM case, we neglect the mechanical couplings and focus on the optics. 
In the symmetric and antisymmetric basis \(\{\a_-,\a_+,\hat d\}\), 
the optical dynamics is governed by
\begin{equation}
\diff*{\begin{pmatrix}
\hat a_-\\
\hat a_+\\
\hat d
\end{pmatrix}}{t}
=
-i
\begin{pmatrix}
\tilde \Delta_- & 0 & 0\\
0 & \tilde \Delta_+ & \lambda_0\\
0 & \lambda_0 & \tilde \Delta_{d,0}
\end{pmatrix}
\begin{pmatrix}
\hat a_-\\
\hat a_+\\
\hat d
\end{pmatrix}
+
\bm D_\RM^{(\pm)},
\label{eq:optical_matrix_RM_pm}
\end{equation}
where
\(\tilde \Delta_\pm
=
\Delta_\pm - i\kappa_a\), 
\(\tilde \Delta_{d,0}
=
\Delta_{d,0} - i\kappa_d\) and $\bm D_\RM^{(\pm)}$ contains the coherent drive and optical noise terms from Eq.~\eqref{eq:QLE_modelB:+-}.

Equation~\eqref{eq:optical_matrix_RM_pm} shows that the antisymmetric mode \(\hat a_-\) is completely decoupled from the Fano mode. The remaining two normal modes arise from the hybridization between the symmetric cavity mode \(\hat a_+\) and the Fano mode \(\hat d\). This two-mode sector has the same structure as the optical normal-mode problem in the TM case, with the replacement \(\a\rightarrow\a_+\) and
\(\Delta_{a,0}\rightarrow\Delta_+\), see Figs.~\subfigref{fig:Model}{b} and \subfigref{fig:Model}{d}. The corresponding complex eigenfrequencies are therefore
\begin{equation}
\Omega^\pm_\RM
=
\bar\Delta_\RM
-
i\bar\kappa
\pm
\sqrt{
\left(
\dD_\RM
-
i\delta\kappa
\right)^2
+
\lambda_0^2
},
\label{eq:Omega_pm_RM}
\end{equation}
where
\begin{equation}
\bar\Delta_\RM
=
\frac{\Delta_++\Delta_{d,0}}{2},
\qquad
\dD_\RM
=
\frac{\Delta_+-\Delta_{d,0}}{2},
\label{eq:Delta_bar_delta_RM}
\end{equation}
and the other symbols remain the same as in the TM case, see Sec.~\ref{sec:modelA:normal modes}.
Thus, although the RM optical sector contains three modes, only two of them participate in the Fano-induced hybridization. The antisymmetric mode \(\hat a_-\) remains a purely FP-like mode, while the symmetric mode \(\hat a_+\) hybridizes with the Fano mode in the same way as the cavity mode \(\hat a\) hybridizes with \(\hat d\) in the TM regime, giving normal modes with detunings $\Delta^\pm_\RM=\Ree(\Omega^\pm_\RM)$ and linewidths $\kappa^\pm_\RM=-\Imm(\Omega^\pm_\RM)$.
This symmetry-selective hybridization is a distinctive feature of the present system and has no analogue in end-mirror Fano setups~\cite{Cernotik2019Jun,WWoe2023,Monsel2024Apr,DL2025Fano,Du2026QST}.

\begin{figure*}[tb]
    \includegraphics[width=\linewidth]{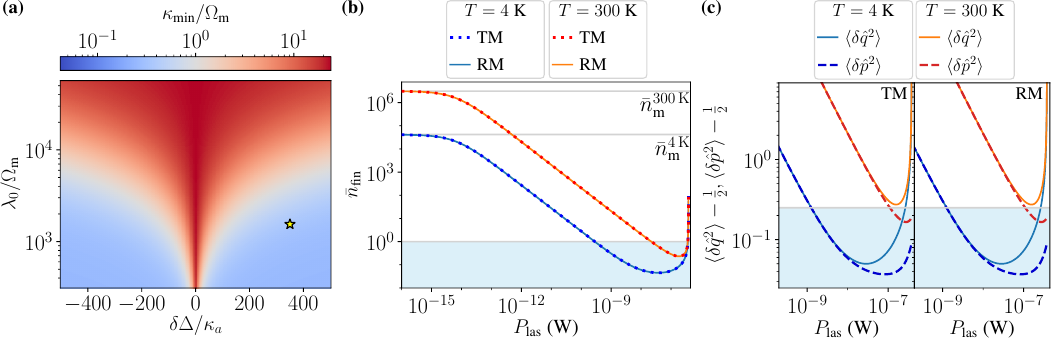}
    \caption{\label{fig:cooling}
        (a) Effective sideband resolution $\kappa_\text{min}/\Om$ as a function of the detuning $\dD$ (see Secs.~\ref{sec:modelA:normal modes} and \ref{sec:modelB:normal modes} for the TM and RM cases respectively) and coherent coupling $\lambda_0$. 
        (b) Steady-state phonon number \(\bar n_{\mathrm{fin}}\) in the mechanical fluctuations as a function of the laser power \(P_{\mathrm{las}}\), for both the TM and the RM cases. In both cases, the narrowest normal mode is driven on its red sideband, \(\Delta^-_\TMRM=\Om\). 
        The horizontal gray lines indicate the thermal phonon numbers at 4 K and 300 K. 
        (c) Mechanical position and momentum fluctuations $\mean{\dq^2}$ and $\mean{\deltp^2}$ as functions of the laser power in the TM and RM cases. 
        Blue shading indicates parameter regimes in which $\nfin < 1$. 
        The yellow star in panel (a) indicates the parameters chosen in panels (b) and (c).
        Unless otherwise indicated, the parameters are $\Om/2\pi = 2$ \si{\mega\hertz}, $Q_\text{m} = 10^8$, $\ka/\Om = 47.7$, $\kd/\Om = 0.25$, $\dD_\TMRM/\ka = 350$, $\lambda_0/2\pi = 3.09$ \si{\giga\hertz}, $\chi/2\pi = 15.9$ \si{\giga\hertz}, $g_{d,0}/2\pi = -182$ \si{\hertz}, and $g_{\lambda, 0}/2\pi=49.6$ \si{\kilo\hertz}; see Sec.~\ref{sec:exp}.
    }
\end{figure*}

\section{Effective sideband resolution and ground-state cooling}\label{sec:cooling}

In contrast to a conventional dielectric MIM system, where the optical linewidth is essentially fixed by the bare cavity decay rate, we have seen in Secs.~\ref{sec:modelA:normal modes} and \ref{sec:modelB:normal modes} that the Fano membrane enables spectral engineering through hybridization with the localized Fano mode. Furthermore, the membrane motion also modulates the cavity–Fano hybridization, introducing an additional optomechanical interaction absent in conventional MIM systems. As we show below, these effects can be harnessed to obtain a sideband-resolved effective optical mode, with a linewidth smaller than the mechanical frequency, and achieve sideband-based ground-state cooling \cite{Aspelmeyer2014Dec, Teufel2011Jul}, even though the bare cavity itself remains in the unresolved-sideband regime.

As discussed in Secs.~\ref{sec:modelA:normal modes} and \ref{sec:modelB:normal modes}, one can first focus on the optical sector and neglect the mechanical coupling. Under this approximation, the relevant hybridization problem is essentially the same in the TM and RM cases. In the latter case, the antisymmetric mode \(\hat a_-\) is decoupled from the rest of the system, as shown in Eq.~\eqref{eq:optical_matrix_RM_pm} and Fig.~\figpanel{fig:Model}{d}, while the remaining modes \(\hat a_+\) and \(\d\) have the same coupling structure as the \(\a\)-\(\d\) subsystem in the TM case. Therefore, the relevant optical normal modes [Eq.~\eqref{eq:Omega_pm_RM}] are described by Eqs.~\eqref{eq:Omega_pm_modelA}, \eqref{eq:NormalMode_def}, and \eqref{DpmTM} with the replacement \(\a\rightarrow \hat a_+\) and \(\Delta_{a,0}\rightarrow\Delta_+\). 
As a consequence, when taking $\dD_\TM = \dD_\RM$, the results presented in this section are independent of whether the membrane is almost transparent or highly reflective. In the following, we will therefore drop the TM/RM subscripts when the distinction is not necessary, such that, e.g., $\dD \equiv \dD_\TMRM$ or $\kappa^\pm \equiv \kappa^\pm_\TMRM$.

Figure~\subfigref{fig:cooling}{a} shows the minimum linewidth of the two optical normal modes\footnote{Note that Eqs.~\eqref{eq:Omega_pm_modelA} and \eqref{eq:Omega_pm_RM} do not give a unique definition of $\Omega^+$ and $\Omega^-$ [and of the associated resonance frequencies and linewidths, Eq.~\eqref{DpmTM}] since they contain the square root of a complex number. To ensure consistency, we take this square root such that $\Omega^+$ and $\Omega^-$ are continuous functions of $\dD$. We furthermore arbitrarily choose to have $\Omega^+$ converge to the Fano membrane mode in the limit $\dD \to -\infty$ while, conversely, $\Omega^-$ converges to the membrane mode in the limit $\dD \to +\infty$. As a result, for the parameters of Fig.~\subfigref{fig:cooling}{a}, $\kappa_{\min} = \kappa^+$ for $\dD < 0$ and $\kappa_{\min} = \kappa^-$ for $\dD > 0$.},
$\kappa_{\min}\equiv \min \left(\kappa^{+},\kappa^{-}\right)$,
 as a function of \(\dD\) and \(\lambda_0\). The blue regions indicate parameter regimes where this hybrid mode becomes sideband-resolved. 
As expected, a narrow optical resonance can be obtained when the Fano mode contributes strongly to one of the hybridized modes. 
$\kappa_{\min}$ is an even function of $\dD$, with $ \kappa^+ = \kappa^- = \bar{\kappa}=(\ka+\kd)/2$ at $\dD = 0$, and we will therefore focus on the case $\dD > 0$ in the following.
In the limit $\dD \gg \lambda_0 \gg \abs{\delta\kappa}$, Eqs.~\eqref{eq:Omega_pm_modelA} and \eqref{eq:Omega_pm_RM} yield
\begin{align}
    \kappa^+ &\simeq \ka - \frac{\lambda_0^2\delta\kappa}{2\dD^2},&
    \kappa^- &\simeq \kd + \frac{\lambda_0^2\delta\kappa}{2\dD^2}.\label{kappa_-} 
\end{align}
A necessary condition for effective sideband resolution is therefore $\kd < \Om$. The remaining contribution to \(\kappa^-\) can be made arbitrarily small by increasing the detuning between the relevant cavity mode and the Fano mode\footnote{We note, however, that the normal-mode description retains only the optical modes included explicitly in Sec.~\ref{SecModel}. Therefore, when $\dD$ becomes comparable to a significant fraction of the free spectral range, neighboring FP resonances may also become relevant. This effect is not captured by the present coupled-mode description.}. Since the Fano-mode frequency \(\omega_d\) is set by the photonic-crystal patterning, this detuning provides a practical design parameter for linewidth engineering. Unlike end-mirror Fano setups~\cite{Cernotik2019Jun,WWoe2023,Monsel2024Apr,DL2025Fano,Du2026QST}, where the linewidth of the Fano mode is essentially limited by its coupling to the external optical reservoir, in the present MIM geometry, it is limited primarily by its intrinsic loss. For the parameter regime considered here (see caption of Fig.~\ref{fig:cooling} and Appendix \ref{app:params}), this intrinsic loss is much smaller than the cavity linewidth, $\kd \ll \ka$, such that $\kappa^-$ is the narrower of the two effective linewidths.

Achieving a favorable sideband-resolution ratio, $\kappa_{\rm min}/\Om \lesssim 1$, is however not sufficient for efficient resolved-sideband cooling. The narrow optical mode must also be accessible from the external drive and must retain a sufficiently strong optomechanical coupling to the mechanical mode. To assess the latter, we compute the single-photon optomechanical couplings associated with the normal modes using Eqs.~\eqref{eq:Omega_pm_modelA} and \eqref{eq:Omega_pm_RM}, assuming that the optomechanics is a small perturbation,
\begin{align}\label{g0-}
    g_{0}^\pm =\,& \frac{1}{\sqrt{2}}\diffp{\Omega^\pm}{q}[q=0] 
    \simeq \frac{g_{d,0}}{2} (1 \mp 1) \pm \frac{\lambda_0}{\dD} g_{\lambda,0},
\end{align}
in the limit $\dD \gg \lambda_0 \gg \abs{\delta\kappa}$. So, we get a coupling of order $g_{d, 0}$ for the narrow ``$-$''-mode and a suppressed but finite coupling for the ``$+$''-mode for the parameters of Fig.~\subfigref{fig:cooling}{b}. However, the effective optomechanical couplings after linearization (Secs.~\ref{sec:lin:modelA} and \ref{sec:lin:modelB}) also depend on the average photon numbers of the modes\footnote{In the RM case, there is also an additional contribution from the three-mode coupling between $\a_-$, $\a_+$ and $\d$ in the effective optomechanical couplings, see Fig.~\subfigref{fig:Model}{d}, Sec.~\ref{sec:lin:modelB}, and Appendix~\ref{app:RMa}.}.
For example, in the absence of intrinsic membrane loss, an extremely narrow optical normal mode can be obtained when the Fano mode is far detuned from the cavity mode. In this limit, however, the hybridization is weak: the narrow mode is only weakly populated by a drive applied through the cavity due to the large detuning. Efficient cooling therefore requires a compromise between reducing the optical linewidth and maintaining enough cavity-Fano hybridization to drive the relevant normal mode, as will be discussed later. 

We evaluate the cooling performance by solving the linearized Langevin equations, including the mechanical mode and the associated noise, as described in Appendixes~\ref{app:TM} and \ref{app:RM}. Figure~\figpanel{fig:cooling}{b} shows the final phonon occupation \(\bar n_{\rm fin}\) as a function of the input laser power in both the TM and RM regimes. In both cases, the parameters are chosen such that the relevant optical mode has a sideband-resolution ratio \(\kappa_{\min}/\Om\simeq 0.35\), and the laser is tuned to the red sideband of that mode, $\Delta^- = \Om$. Red and orange curves correspond to room temperature, \(T=300\,\si{\kelvin}\), while blue curves correspond to cryogenic temperature, \(T=4\,\si{\kelvin}\). The cooling curves evidence that the narrow normal mode can reduce the mechanical occupation well below unity over a finite range of input powers, in both the TM and RM limits. The other requirement to achieve ground-state cooling, besides $\nfin < 1$, is \emph{energy equipartition}, namely $\langle\dq^2\rangle\simeq  \langle\deltp^2\rangle$. We check this second condition in Fig.~\figpanel{fig:cooling}{c}, demonstrating that, at $T = 4$ K, the mechanical mode is indeed cooled close to its ground state. This is, however, not fully the case at room temperature, where the laser power needed to reach $\nfin < 1$ is too high and already breaks equipartition. The mechanics is nevertheless quite close to its ground state, and operating at lower temperature, e.g. with liquid nitrogen cooling, or 
increasing the quality factor $Q_{\rm m}$ of the mechanical resonator would reduce the thermal load and facilitate ground-state cooling.
These results show  that ground-state cooling is achievable without requiring the bare cavity to be sideband resolved in both the TM and RM regimes. This demonstrates the key advantage of introducing a Fano membrane into a MIM architecture. We discuss the experimental feasibility of this sideband-cooling scheme in Sec.~\ref{sec:exp}.

\begin{figure}[tb]
    \includegraphics[width=\linewidth]{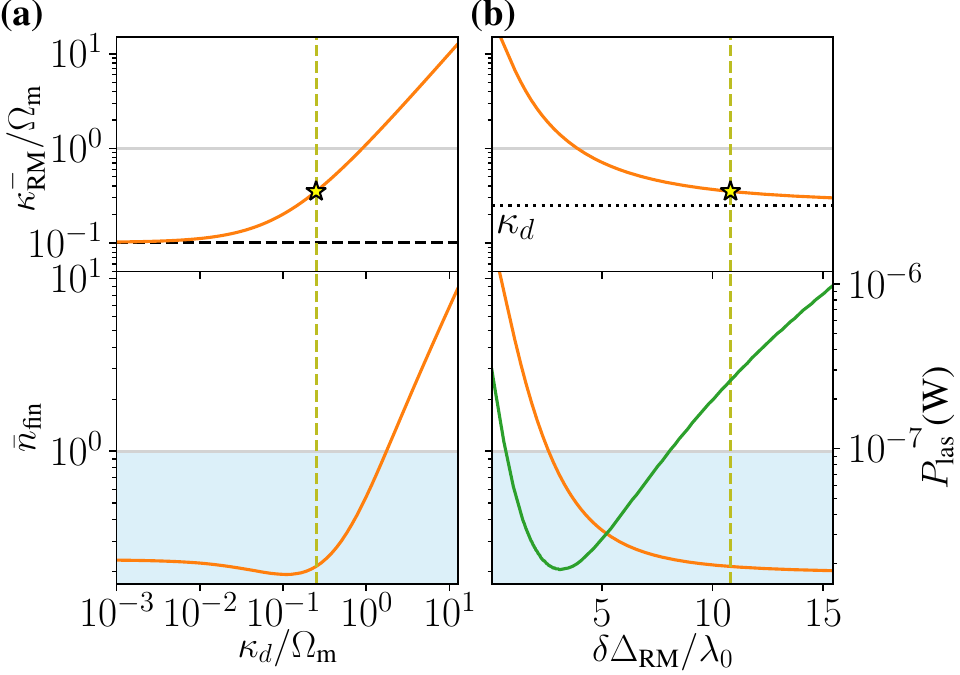}
    \caption{\label{fig:cooling2}
        Effective sideband resolution (top) and the corresponding steady-state phonon number (bottom), minimized over $\Delta^-_\RM$ and $P_\las$, as functions of (a) the intrinsic Fano-mode decay rate $\kd$ and (b) the detuning $\dD_\RM$, varied by changing $\omd$.
        In panel (a), the dashed black line indicates the effective sideband resolution in the limit $\abs{\delta\Delta_\RM} \gg \lambda_0$ and $\kd \to 0$, $\kappa_\text{lim} = \lambda_0^2 \ka / (4\delta \Delta_\RM^2)$ [Eq.~\eqref{kappa_-}]. 
        In the bottom of panel (b), the laser power at which the minimum $\nfin$ is reached (green curve) is shown on the right axis.
        The blue shaded areas indicate parameter regimes in which $\nfin < 1$.
        Unless otherwise indicated, the parameters are the same as for the RM case in Fig.~\subfigref{fig:cooling}{b}, at $T = 300$ K. The yellow stars indicate the exact parameters of Fig.~\subfigref{fig:cooling}{b}.
    }
\end{figure}

Figure~\figpanel{fig:cooling2}{a} further illustrates the role of the intrinsic Fano-mode loss inferred from Eq.~\eqref{kappa_-}. As \(\kappa_d\) increases, the effective sideband resolution deteriorates. As a result, the optimized cooling performance, that is the phonon number $\nfin$ minimized over the detuning $\Delta^-$ and the laser power $P_\las$, worsens, eventually making ground-state cooling impossible. In the opposite limit of a lossless membrane, $\kd \to 0$, $\kappa^-$ reaches the finite value $\lambda_0^2 \ka / (4\delta \Delta^2)$ (black dashed line) due to the hybridization with the cavity mode. Figure~\ref{fig:cooling2} only shows the RM case, but, like in Fig.~\ref{fig:cooling}, the TM case gives equivalent results.

Finally, we highlight the trade-off between the sideband resolution ratio and the cooling efficiency due to the hybridization between the broad cavity mode and the narrow Fano mode in Fig.~\subfigref{fig:cooling2}{b}. The top panel shows how the sideband resolution ratio decreases with the detuning $\dD_\RM$ between the $a_+$ mode and the $d$ mode [Eq.~\eqref{kappa_-}], allowing for reaching lower phonon numbers $\nfin$ (orange curve in the bottom panel). However, once the system enters the sideband-resolved regime, increasing $\dD_\RM$ comes at the cost of requiring higher laser powers to reach the minimum phonon number $\nfin$, as shown in the bottom panel of Fig.~\subfigref{fig:cooling2}{b} (green curve, right axis); see Appendix \ref{app:efficiency} for more details.

\section{Experimental feasibility}\label{sec:exp}

The parameters required to achieve the sufficiently small values of $\kappa_{\min}$ in Sec.~\ref{sec:cooling} to enter the effective sideband-resolved regime are within reach of state-of-the-art experiments. 
However, this is not straightforward to assess since the parameters of the coupled-mode models presented in Sec.~\ref{SecModel} do not relate in a simple way to the geometry and material choices of a specific device. Conversely, the well-established transfer-matrix approach \cite{Klein1986, Deutsch1995Aug, Xuereb2009May, Cernotik2019Jun} establishes a direct connection between the cavity length, the reflectivities of the mirrors and membrane, and the optical properties of the setup. In particular, the identical end mirrors are characterized by their polarizability $\zN$, and the direct optical response of the unpatterned membrane (without the Fano mode) is characterized by the polarizability $\zD$. A low polarizability, $\abs{\zD} \ll 1$, corresponds to the TM case, while a large polarizability, $\abs{\zD} \gg 1$, corresponds to the RM case. Finally, the relevant photonic-crystal mode of the membrane can be described by a Fano resonance of frequency $\omd$ with a Fano linewidth\footnote{Not to be confused with the loss rate $\kd$ that models internal losses in the membrane, such as absorption losses.} $\gF$ \cite{Fan2003Mar, Cernotik2019Jun}. We have used this transfer-matrix approach to choose realistic parameters for Figs.~\ref{fig:cooling} and \ref{fig:cooling2}. We summarize briefly here how the parameters of the transfer matrix model are linked to those of the coupled-mode models, while detailed explanations are provided in Appendix \ref{app:transfer matrix}.

In the TM case, $\oma$ is the resonance of a FP cavity of length $2L$, see Fig.~\subfigref{fig:Model}{a}, which is closest to $\omd$, but with a small frequency shift because the end mirrors are not perfectly reflective and the membrane is not fully transparent. The total loss rate of the cavity is given by the usual expression for a FP cavity of length $2L$, $\ka = (c/4L)/\zN^2$  \cite{Cernotik2019Jun}, where $c/4L$ is the free spectral range and $c$ the speed of light. The coupling strength 
\begin{equation}\label{lbd:TM}
    \lambda_0 = 2\sqrt{\frac{c}{4L}\gF}
\end{equation}
comes from the overlap between the relevant cavity mode (free spectral range $c/4L$) and the membrane mode (Fano linewidth $\gF$), see Appendix \ref{app:lambda} for details.

In the RM case, the setup consists of two identical FP cavities of length $L$ coupled together. Accordingly, $\omega_0$ corresponds to the resonance (closest to $\omd$) of a FP cavity of length $L$, whose two mirrors have polarizabilities $\zN$ and $\zD$, respectively. $\zD$ is the direct polarizability of the membrane (in the absence of photonic-crystal patterning \cite{Fan2003Mar}), and the coupling between the left and right cavities is given by
\begin{equation}\label{chi:RM}
    \chi = \frac{c}{2L} \frac{1}{\zD},
\end{equation}
where $c/2L$ is the free spectral range of the cavities.
Since the cavities lose light to the environment only through one mirror, we find that their loss rate is $\ka = (c/2L)/2\zN^2$, which is equal to the total optical loss rate in the TM case. Furthermore, each cavity mode overlaps with the membrane mode, giving $\lambda_{\L,0} = \lambda_{\R,0} = \sqrt{(c/2L) \gF}$. The factor 2 difference compared to the TM case [Eq.~\eqref{lbd:TM}] comes from the fact that each cavity mode overlaps with the membrane mode on only one side. Then, the coupling between $\a_+$ and the membrane mode $\d$ gives a factor $\sqrt{2}$ enhancement, such that the coupling $\lambda_0$ has the same value in both cases. In the RM case, there is a nonnegligible dispersive optomechanical coupling of each cavity mode coming from the displacement of the membrane, such that $g_{a, 0} = (\omega_0/L)\xzpf$ \cite{Aspelmeyer2014Dec}, where $\xzpf$ is the zero-point fluctuation amplitude of the membrane.  

Based on this matching between the transfer-matrix approach and the coupled-mode models, we can assess how experimentally realistic the parameters from Figs.~\ref{fig:cooling} and \ref{fig:cooling2} are. First, they correspond to a distance $L = 50$ \si{\micro\meter} between the membrane and the end mirrors, which is relatively short but achievable  \cite{SaarinenOptica2023}. Short distances $L$ are desirable to enhance the coupling strength $\lambda_0$, see Eq.~\eqref{lbd:TM}. The polarizability of the end mirrors is $\zN = 50$, but lower polarizabilities, that is higher loss rates $\ka$, can be compensated by increasing the detuning $\dD_\TMRM$ [Eq.~\eqref{kappa_-}]. The direct polarizability of the membrane is $\zD = 0.001$ in the TM case and $\zD = 30$ in the RM case, but polarizabilities down to $\zD \gtrsim 1$ are sufficient in the latter case; see Appendix \ref{app:lambda}. Second, the laser frequencies we consider are in the telecom wavelength regime (1480 nm to 1620 nm), like in Refs.~\cite{WWoe2023, SaarinenOptica2023}, see also Appendix \ref{app:params}, and the mechanical parameters are realistic for two-dimensional suspended phononic-crystal membranes, for instance made of SiN \cite{SaarinenOptica2023}, InGaP \cite{Manjeshwar2023Jun}, or AlN \cite{CiersAPL2025}. 
Then, the resonance frequency $\omd$ and Fano linewidth $\gF$ of the membrane can be engineered by changing the photonic-crystal pattern (lattice constant, hole radius) \cite{Fan2003Mar, Peralle2024Jan, WWoe2023}, making it possible to choose appropriate values of $\dD$ and $\lambda_0$. 
The most stringent requirement to reach the effective sideband-resolved regime is to have $\kd < \Om$ [Eq.~\eqref{kappa_-}], which requires a membrane with minimal absorption and scattering losses across the wavelength range relevant to the device operation.

The optomechanical coupling strengths of the Fano mode are based on the corresponding dispersive and dissipative couplings measured in the experimental device of Ref.~\cite{WWoe2023}, but are reduced by a factor $10^4$ to ensure operation in the linear regime; see Appendixes \ref{app:om couplings} and \ref{app:params}. The dispersive coupling $g_{a,0}$ was obtained by assuming $\xzpf = \sqrt{\hbar/2m\Om}= 1$ \si{\femto\meter}, based on a motional mass $m$ of the order of a nanogram \cite{CiersAPL2025}. Finally, even though we are operating in the perfectly symmetric configuration, which is typically an operating point to access quadratic optomechanics \cite{Jayich2008Sep, Sankey2010Sep}, in the setup we consider, quadratic optomechanical effects are negligible compared to the linear optomechanical effects from the photonic-crystal membrane. Indeed, the experimentally reported values reach up to $\sim 30\, \si{\mega\hertz/\nano\meter^2}$ \cite{Sankey2010Sep, Lee2015Feb}, which, for $\xzpf = 1$ \si{\femto\meter}, results in a single-photon quadratic coupling strength below $10^{-4}\,\si{\hertz}$. This is negligible compared with the \si{\kilo\hertz}-scale values of $g_{d,0}$ and $g_{\lambda, 0}$, even though the latter have already been considerably reduced compared to the experimentally reported values \cite{WWoe2023}. The large single-photon optomechanical coupling $g_{d, 0}$ reported in Ref.~\cite{WWoe2023} would even enable access to the strong-coupling regime, where $\kappa_{\min}$ approaches the effective single-photon optomechanical coupling, paving the way for quantum nonlinear optomechanics.

\section{Conclusion and discussion}

In summary, we have studied cavity optomechanics with a photonic-crystal Fano membrane placed at the center of a FP cavity. The frequency-dependent optical response of the membrane adds a localized Fano mode to the conventional MIM description and introduces optomechanical couplings beyond the standard displacement-induced shift of the cavity resonances. In particular, the membrane motion can modify both the Fano-mode resonance and the coherent coupling between the Fano and cavity modes. This leads to distinct effective descriptions in the TM and RM limits. In the RM case, the symmetry of the cavity modes is especially important: only the symmetric superposition hybridizes with the Fano mode, whereas the antisymmetric mode remains uncoupled from it. Compared with setups where the Fano membrane serves as a cavity end mirror, the Fano mode in this MIM configuration is subject only to weak intrinsic optical losses, rather than direct coupling to external optical ports. This provides a route to substantially reduce the linewidth of the relevant optical normal mode.

Our analysis shows that this hybridization is not simply the appearance of a narrow optical resonance. For optomechanical cooling, the narrow optical normal mode, which is relevant for control over the mechanical mode, must be efficiently driven and couple appreciably to the mechanical motion. The resulting trade-off between linewidth reduction and optical accessibility determines the relevant operating regime. Within the parameter range considered here, this normal mode can become effectively sideband resolved even when the bare cavity is not, enabling ground-state cooling of the membrane motion with parameters achievable in state-of-the-art devices. 

The transfer-matrix method provides a direct connection between these effective mode descriptions and the frequency-dependent optical response of a realistic photonic-crystal membrane. Through this connection, the parameters of the coupled-mode model can be estimated from the device geometry together with independent characterization of the optical and mechanical components. This framework is therefore useful for identifying how the membrane patterning, reflectivity, and cavity geometry determine the effective parameters governing the optomechanical dynamics. 
In view of this, a Fano membrane in the middle provides additional degrees of freedom for tailoring both the relevant optical linewidths and optomechanical interactions within a single device. 

Finally, the large single-photon optomechanical coupling strengths reported for Fano modes in Ref.~\cite{WWoe2023}, albeit in a membrane-at-the-end geometry, make this platform particularly promising for accessing the quantum nonlinear regime of optomechanics. This opens the possibility of combining linewidth engineering with enhanced cooperativity and nonclassical dynamics, paving the way for applications such as quantum sensing and the preparation of mechanical resonators in nonclassical quantum states \cite{Hauer2023prl, Du2026QST}. Future extensions to asymmetric cavities \cite{Dumont2019Sep}, multiple Fano membranes, and regimes of stronger optomechanical nonlinearity may reveal even more favorable configurations for these applications.

\acknowledgments

We thank Alexander Jung, Anastasiia Ciers, and Ludovico Tesser for useful discussions. We acknowledge financial support from the Knut och Alice Wallenberg stiftelse through Project No. 2022.0090, as well as through the Wallenberg Scholar grant (W.W.). 

\section*{Data availability statement}

The data that support the findings of this article are openly available \cite{Zenodo}.

\appendix

\section{Transfer-matrix approach: bridging the gap between low and high membrane reflectivities}\label{app:transfer matrix}

In this Appendix, we model the optical response of a MIM setup in any membrane reflectivity regime using a classical transfer matrix approach \cite{Klein1986, Deutsch1995Aug, Xuereb2009May}. Comparing the results to those of the coupled-mode models in the TM and RM limits (see Sec.~\ref{SecModel}), we establish the relations between the parameters of the different models. This is particularly useful from an experimental perspective since the transfer-matrix model parameters are easier to estimate from the geometry and materials of the device than the coupled-mode model parameters, as discussed in Sec.~\ref{sec:exp}.

First, we introduce the transfer-matrix approach in Appendix \ref{app:tm} and summarize the parameter matching with the coupled-mode models in Appendix \ref{app:comp}. The other sections give technical details about how the relations between parameters were established: Appendix \ref{app:FP} presents the method used to identify resonances in the case of a standard Fabry-Pérot cavity, Appendix~\ref{app:no PhC} studies the resonances without the photonic crystal, and Appendix~\ref{app:lambda} considers the full setup to determine the coupling $\lambda_0$. Appendix~\ref{app:om couplings} finally discusses the consequences for the optomechanical couplings.

\subsection{Transfer-matrix model}\label{app:tm}
The setup, as depicted in Fig.~\subfigref{fig:Model}{a}, can be modeled as three successive scatterers \cite{Klein1986, Deutsch1995Aug}: the left mirror, the photonic-crystal (Fano) membrane and the right mirror. In between these scatterers, the electromagnetic field propagates freely, and, like previously, we assume that the left and right mirrors are identical, highly reflective, and frequency independent. The electromagnetic fields on the left and right sides of the setup are therefore related by the transfer matrix \cite{Klein1986, Deutsch1995Aug, Xuereb2009May} $\bM = \bM_\N \bM_\text{prop} \bM_\F \bM_\text{prop}\bM_\N$, with
\begin{align}
    \bM_\text{N} &= \frac{1}{t_\text{N}}\begin{pmatrix}
        1 & -r_\text{N} \\ r_\text{N} & t^2_\text{N} - r^2_\text{N}
    \end{pmatrix},\label{M_N} \\
    \bM_\text{F} &= \frac{1}{t_\text{F}(\omega)}\begin{pmatrix}
        1 & -r_\text{F}(\omega) \\ r_\text{F}(\omega) & t^2_\text{F}(\omega) - r^2_\text{F}(\omega)
    \end{pmatrix},\label{M_F}\\
    \bM_\text{prop} &= \begin{pmatrix}
        \e^{-i\frac{\omega L}{c}} & 0 \\ 0 & \e^{i\frac{\omega L}{c}}
    \end{pmatrix}.\label{M_prop}
\end{align}
$L$ denotes the distance between the membrane and each of the end mirrors and $c$ is the speed of light. $t_\mathrm{N/F}$ and $r_\mathrm{N/F}$ are the transmission and reflection coefficients of the scatterer. For the end mirrors, they are constant and given by \cite{Deutsch1995Aug}
\begin{align}\label{t_N, r_N}
    t_\N & = \frac{1}{1 - i\zN} = \cos\thN\e^{i\thN}, \\ \nonumber
    r_\N & = \frac{i\zN}{1 - i\zN} = i\sin\thN\e^{i\thN},
\end{align}
where $\zN = \tan\thN$ is the polarizability of the end mirrors. Without loss of generality, we choose the phase shift upon transmission, $\thN$, such that $\sin\thN > 0$, namely by taking $\thN = \arctan(\zN)$ for $\zN > 0$ and $\thN = \pi + \arctan(\zN)$ for $\zN < 0$. 
The membrane can also be described in the same way, but with frequency-dependent transmission and reflection coefficients \cite{Fan2003Mar, Cernotik2019Jun},
\begin{align}
    t_\F(\omega) &= \frac{-r_\D \gF + i t_\D( \omF - \omega)}{\gF + i(\omF - \omega)},\nonumber\\
    r_\F(\omega) &=  \frac{- t_\D \gF +i r_\D(\omF - \omega)}{\gF + i(\omF - \omega)} \label{t r Fano}.
\end{align}
Here we denote by $\omF$ and $\gF$ the frequency and linewidth of the Fano resonance. $t_\D = \cos\thD\e^{i\thD}$ and $r_\D =i \sin\thD\e^{i\thD}$ are the transmission and reflection coefficients for the direct propagation process through the Fano membrane \cite{Fan2003Mar}, with polarizability $\zD = \tan\thD$, where we have chosen $\thD$ such that $\sin\thD > 0$. This model is particularly practical from an experimental perspective as $t_\N$, $r_\N$, $r_\D$, $t_\D$, $\omF$, and $\gF$ can be estimated based on the refractive indices, geometry, and fits to either finite-element simulations or experimental data from the isolated mirror or membrane. Note that Eq.~\eqref{t r Fano} describes a lossless Fano membrane, for which $|t_F(\omega)|^2+|r_F(\omega)|^2=1$. A lossy extension can be obtained by modifying the transmission and reflection amplitudes such that the energy relation becomes~\cite{Mitra2024OE}
\begin{equation}
|t_F^{\rm loss}(\omega)|^2+|r_F^{\rm loss}(\omega)|^2+\frac{c_{\rm loss}^2}{(\omega-\omega_F)^2+\gamma_F^2}=1,
\end{equation}
where $c_{\rm loss}$ characterizes the strength of the loss channel. For simplicity, in this Appendix, we restrict the transfer-matrix analysis to the lossless case, where $\zN$ and $\zD$ are real numbers and $c_{\rm loss}=0$.

From this transfer matrix approach, we obtain the mean-field transmission amplitude \cite{Deutsch1995Aug, Xuereb2009May}, 
\begin{align}\label{tM}
    t_\tm(\omega) &= \frac{1}{[\bM]_{11}} \\\nonumber
    &=\frac{{t_\F(\omega)} t_\N^2\e^{i \frac{2\omega L}{c}}}{ {\left({r_\F(\omega)} {r_\N}\e^{i \frac{2\omega L}{c}}  - 1\right)^2 - r_\N^2 t_\F(\omega)^2\e^{i \frac{4\omega L}{c}}}}
\end{align}
Our goal is to now interpolate between $t_\TM$ [Eq.~\eqref{eq:transmission_modelA}] and $t_\RM$ [Eq.~\eqref{eq:transmission_modelB}], identifying the relations between the coupled-mode model parameters ($\oma, \omd, \ka, \kd, \lambda_0, \chi$) and the transfer-matrix model parameters ($L, \zN, \zD, \omF, \gF$). 

\subsection{Relation to the coupled-mode models}\label{app:comp}

To make the comparison with the coupled-mode models, we keep  $L$, $\zN$, $\omF$, and $\gF$ fixed and assume we can vary the polarizability $\zD$ alone. Since $\abs{\zD}^2 = R_\D / T_\D$, where $T_\D = \abs{t_\D}^2$ (resp. $R_\D= \abs{r_\D}^2$) is the energy transmission (resp. reflection) coefficient of the unpatterned membrane, Sec.~\ref{sec:modelA} corresponds to $\abs{\zD} \ll 1$ and Sec.~\ref{sec:modelB} to $|\zD| \gg 1$. We have summarized the results of the parameter matching in Table~\ref{tab:param matching}, and we detail how we obtained those results in Appendixes \ref{app:FP} to \ref{app:om couplings}.
The frequency of the resonant feature in the membrane corresponds to the localized optical mode (the Fano mode) from Secs.~\ref{sec:modelA} and \ref{sec:modelB}, such that $\omF = \omd$. We have not included any dissipation in $M_\text{F}$ [Eq.~\eqref{M_F}], which means that we should compare to the case $\kd = 0$. 


\begin{table}[b]
    \renewcommand{\arraystretch}{1.5}
    \setlength{\tabcolsep}{4pt}
    \begin{tabular}{llc}
        \hline
        \hline
        Parameter& Expression & Model \\
        \hline
        $\oma$ & $\pi n_\TM\FSR+ \left(\pi/2 -\thN - 2\thD\right)\FSR$ & TM\\
        $\omega_{a_\L} \!= \omega_{a_\R} \!\equiv \omega_0$ & $2\pi n_\RM \FSR + \left(\pi - \thN - \thD\right)\FSR$ & RM\\
        $\omd$ & $\omF$ & both\\
        $\ka$ & $-\FSR \ln(\sin\thN) \simeq \FSR/2\zN^2$ & both\\
        $\kd$ & $0$ & both \\
        $\lambda_0$ & $\sqrt{2\FSR\gF}$ & both \\
        $\chi$ & $(\pi/2-\thD)\FSR \simeq \FSR/\zD$ & RM \\
        $g_{a, 0}$ & $\omega_0\xzpf/L$ & RM \\
        $g_{\lambda, 0}$ & $g_{\gamma, 0} \lambda_0/2\gF$ & both \\
        \hline\hline 
    \end{tabular}
    \caption{\label{tab:param matching} 
        Parameter expressions of the coupled-mode model as functions of the transfer-matrix model parameters $L, \thN, \thD, \omF, \gF$, and the polarizabilities $\zN = \tan\thN$ and $\zD = \tan\thD$. $\FSR = c/2L$ is the free spectral range of a FP cavity of length $L$, and $n_\TMRM$ the integer giving the value of $\oma/\omega_0$ closest to $\omd$. We denote by $\xzpf$ the mechanical zero-point fluctuation amplitude and $g_{\gamma, 0}$ the dissipative optomechanical coupling of the Fano resonance (see Appendix~\ref{app:om couplings}).
    }
\end{table}

\subsubsection{Transparent membrane case}
For a fully transparent membrane ($\zD \to 0$), we expect the cavity modes to correspond to resonances of a Fabry-Pérot cavity of length $2L$ with identical end mirrors (see Appendix \ref{app:FP}),
\begin{equation}
    \omega_n = n\frac{\pi c}{2L} + \left(\pi - 2\thN\right)\frac{c}{4L}.
\end{equation}
In the TM case, we consider an almost transparent membrane, with $\abs{\zD} \ll 1$, but finite. As a consequence, the membrane without photonic-crystal patterning, i.e. $\gF \to 0$, introduces an additional phase shift due to the small polarizability $\zD \simeq \thD$, giving the expression in Table~\ref{tab:param matching}, see Appendix \ref{app:no PhC} for details. The relevant resonance for the coupled-mode model is the one with the integer $n = n_\TM$ such that $\oma$ is the FP resonance closest to $\omd$. 
We assume that $\omd$ is close enough to one of the FP resonances and the optical quality factor is large enough to be able to neglect all the other resonances, namely $\abs{\oma - \omd}$ and $\ka$ much smaller than the free spectral range. The cavity decay rates are given by the usual expression for a FP cavity of length $2L$ with highly reflective identical mirrors, see Appendix~\ref{app:FP},
\begin{equation}
    \kL = \kR = \frac{\ka}{2} \simeq \frac{c}{4L} \frac{1}{2 \zN^2}.
\end{equation}
The expression of the coupling strength $\lambda_0$ in Table~\ref{tab:param matching} is the same as Eq.~\eqref{lbd:TM}, since $\FSR = c/2L$ is twice the free spectral range of a FP cavity of length $2L$; see derivation in Appendix~ \ref{app:lambda}.

\subsubsection{Reflective membrane case}

In the opposite limit of a highly reflective membrane, there are two identical FP cavities of length $L$ where one of the end mirrors is the membrane, of polarizability $\abs{\zD} \gg 1$, without photonic-crystal patterning \cite{Cernotik2019Jun}, giving the expression of $\omega_0$ in Table~\ref{tab:param matching}, see Appendix \ref{app:no PhC} for details. The two cavities are each coupled to the environment on one side only, since they are coupled to each other through the membrane, such that their individual loss rate is equal to the total loss rate of a FP cavity of length $2L$ with identical end-mirrors of polarizability $\zN$. 
Since the membrane's reflectivity is high, but not perfect, it introduces a coupling $\chi$ between the left and right cavity modes [Eq.~\eqref{chi:RM}] that is inversely proportional to the membrane polarizability; see Appendix \ref{app:no PhC}. As explained in Sec.~\ref{sec:exp}, the coupling strength $\lambda_0$ has the same expression in both the TM and RM cases; see derivation in Appendix~ \ref{app:lambda}.

\subsection{Resonances of a standard Fabry-Pérot cavity}\label{app:FP}

In this section, we present our method for finding the optical resonances in the textbook case of a FP cavity consisting of a left and right mirror, separated by a distance $L$, with frequency-independent transmission and reflection coefficients $t_{\L/\R}, r_{\L/\R}$ [Eq.~\eqref{t_N, r_N}].  

The FP cavity can be described by the transfer matrix $\bM_\text{FP} = \bM_\L \bM_\text{prop}\bM_\R$ [Eqs.~\eqref{M_N} and \eqref{M_prop}], such that we get the transmission coefficient of the cavity
\begin{align}
    t_\text{FP}(\omega) &= \frac{1}{[\bM_\text{FP}]_{11}} = \frac{t_\L t_\R\e^{i\frac{\omega L}{c}}}{1 - r_\L r_\R \e^{i2\frac{\omega L}{c}}}.
\end{align}
The resonances are given by the poles $\tilde{\omega} = \omega - i\gamma $ of the transmission $t_\text{FP}$, where $\omega$ is the resonance frequency and $\gamma$ the linewidth. They are the solutions of 
\begin{equation}
    \e^{i\tilde{\omega}/{\FSR}} = \frac{1}{r_\L r_\R} = \frac{-\e^{-i(\theta_\L + \theta_\R)}}{\sin\theta_\L \sin\theta_\R}, \label{poleFP}
\end{equation}
where $\FSR = c/2L$ and we have expressed $r_\ell$ as functions of the angle $\theta_\ell$ [Eq.~\eqref{t_N, r_N}], for $\ell = \L,\R$, taking, as previously, $\sin\theta_\ell >0$ .

First, by taking the absolute value of Eq.~\eqref{poleFP}, we find
\begin{equation}
    \gamma = -\FSR\ln(\sin\theta_\L)-\FSR\ln(\sin\theta_\R),
\end{equation}
which can straightforwardly be decomposed into contributions from the left and right:
\begin{align}
    \gamma_\ell &= -\FSR\ln(\sin\theta_\ell) 
    \simeq \frac{\FSR}{2\zeta_\ell^2},
\end{align}
in the limit $\abs{\zeta_\ell} \gg 1$ (good mirror), using Eq.~\eqref{t_N, r_N}.

Second, from the phase in Eq.~\eqref{poleFP}, we get the resonance frequencies
\begin{equation}\label{freq_res}
    \omega_n^\text{FP} = n2\pi\FSR + \left(\pi - \theta_\L  - \theta_\R \right)\FSR, 
\end{equation}
where $n$ is an integer.

\subsection{Cavity resonances without the photonic crystal}\label{app:no PhC}

In this section, we explain in detail how we found the expressions of $\oma, \omega_0, \ka$ and $\chi$ (see Table~\ref{tab:param matching}) by considering the membrane in the middle setup in the limit $\gF \to 0$. In that limit, Eq.~\eqref{t r Fano} becomes $t_\F(\omega) = t_\D$ and $r_\F(\omega) = r_\D$, amounting to having a membrane without photonic crystal. Therefore, the amplitude transmission of the setup [Eq.~\eqref{tM}] becomes
\begin{align}\label{tTM0}
    t_\tm^{0}(\omega)
    &=\frac{{t_\D} t_\N^2\e^{i\frac{\omega}{\FSR}}}{ {\left[(r_\D - t_\D) {r_\N}\e^{ i\frac{\omega}{\FSR}} - 1\right]\!\left[(r_\D + t_\D) {r_\N}\e^{i\frac{\omega}{\FSR}} - 1  \right]}}\nonumber\\
    &=\frac{{t_\D} t_\N^2\e^{i\frac{\omega}{\FSR}}}{ {\left[-{r_\N}\e^{ i\frac{\omega}{\FSR}} - 1\right]\!\left[{r_\N}\e^{i2\thD}\e^{i\frac{\omega}{\FSR}} - 1  \right]}}.
\end{align}
We have now two equations for the poles $\tilde{\omega}^0_\pm = \omega^0_\pm - i \gamma^0_\pm$,
\begin{align}
    \e^{i\tilde{\omega}^0_\pm/\FSR} = \frac{\mp i \e^{-i[(1 \pm 1)\thD + \thN]}}{\sin\theta_\N}, \label{pole2c}
\end{align}
Proceeding like in Appendix \ref{app:FP}, we find the linewidths
\begin{equation}
    \gamma^0_\pm = -\FSR \ln(\sin\thN) \simeq \frac{\FSR}{2\zN^2} \equiv\gamma^0, \label{k0}
\end{equation}
since $\zN \gg 1$, and the two families of resonance frequencies
\begin{subequations}\allowdisplaybreaks\label{w0pm}
    \begin{align}
        \omega^0_{n, +} &= n2\pi\FSR + \left(\frac{3\pi}{2} - 2\thD - \thN\right)\FSR,\label{w0p}  \\
        \omega^0_{n, -} &= n2\pi\FSR + \left(\frac{\pi}{2} - \thN\right)\FSR,\label{w0m}
    \end{align}
\end{subequations}
where $n$ is an integer.

\subsubsection{Transparent membrane limit}\label{app:no PhC:TM}
In the fully transparent membrane limit, $\zD \to 0$, that is $\thD \to 0$, the two resonance families from Eq.~\eqref{w0pm} have merged into a single one,
\begin{equation}\label{wres_TM}
    \omega_{n'}^{0,\TM}  =  n'2\pi\frac{\FSR}{2} + \left(\pi - 2\thN\right)\frac{\FSR}{2},
\end{equation}
where $n'$ is an integer and $\omega_{2n}^{0,\TM} = \omega^{0,\TM}_{n, -}$ and $\omega_{2n+1}^{0,\TM} = \omega^{0,\TM}_{n, +}$.
This is what would be expected from a standard FP cavity of length $2L$ with $\theta_\L = \theta_\R = \thN$ [Eq.~\eqref{freq_res}].

To get a better match in the small but finite polarizability limit, we take $\oma = \omega^0_{n, +}$, which gives the expression from Table~\ref{tab:param matching}. We can neglect $\omega^0_{n, -}$ in the coupled mode model because it is separated from $\oma$ by $\pi\FSR\gg \gamma^0$, and therefore will not be populated when driving the cavity close to $\oma$. The linewidth from Eq.~\eqref{k0} is unchanged, but can be rewritten as $\gamma^0 = (\FSR/2)/\zN^2 \equiv \ka$, which is, again, consistent with a FP cavity of length $2L$ and identical mirrors of polarizability $\zN$. 

\subsubsection{Reflective membrane limit}\label{app:no PhC:RM}

We now consider the opposite limit, $\zD \to \infty$. Then, $\thD \to \pi/2 $ and, therefore, the two families of resonance frequencies from Eq.~\eqref{w0pm} become identical, 
\begin{equation}\label{wres_RM}
    \omega_n^{0,\RM}  =  n2\pi\FSR + \left(\frac{\pi}{2} - \thN\right)\FSR,
\end{equation}
which is consistent with having two uncoupled FP cavities of length $L$, like in Eq.~\eqref{freq_res}, but with one perfectly reflective mirror inducing a phase shift $\pi/2$. We also note that, for any value of the reflectivity of the membrane, $\omega_n^{0,\RM} = \omega_{2n}^{0,\TM} = \omega^0_{n, -}$ [Eqs.~\eqref{wres_RM}, \eqref{wres_TM}, and \eqref{w0m}] is always a resonance of the system.

For a large but finite value of $\abs{\zD}$, the two families of resonance frequencies [Eq.~\eqref{w0pm}] become distinct, with $\thD \simeq \arccos(1/\zD) \simeq \pi/2 - 1/\zD$. As a result, $\omega^0_{n, +} \simeq \omega_n^{0,\RM} + 2\FSR/\zD $ and $\omega^0_{n, -} = \omega_n^{0,\RM}$  are close to each other, separated by 
\begin{equation}
    \Delta\omega^0 = (\pi - 2\thD)\FSR \simeq \frac{2\FSR}{\zD}.\label{dw0}
\end{equation}
In the coupled mode model from Sec.~\ref{sec:modelB}, when the membrane mode is decoupled ($\lambda_0\to 0$), the optical normal modes are at $\omega_0 \pm \chi$ and are therefore separated by $2\chi$; see Sec.~\ref{sec:modelB:normal modes}. Matching this result with Eq.~\eqref{dw0}, we get $\omega_0 \pm \chi = \omega_{n, \pm}^0$, with
\begin{align}
    \omega_0 &= \frac{\omega^0_{n, +} + \omega^0_{n, -}}{2}\label{w0} \\\nonumber&= n2\pi\FSR + \left(\pi - \thN-\thD\right)\FSR,\\
    \chi &=  \left(\frac{\pi}{2}-\thD\right)\FSR\simeq \frac{\FSR}{\zD}, \label{chi}
\end{align}
where $n$ is the integer giving the frequencies $\omega_{n,\pm}^0$ closest to $\omF$, as shown in the next section, Appendix~\ref{app:lambda}. Therefore, $\omega_0$ corresponds to a resonance of a FP cavity of length $L$ with one mirror of polarizability $\zN$ and one of polarizability $\zD$ [Eq.~\eqref{freq_res}].

\subsection{Determination of the coupling $\lambda_0$}\label{app:lambda}
The next step is to also take into account the photonic crystal patterned on the membrane, that is to look at finite values of $\gF$ and $\lambda_0$ to determine how they are related. From the transmission [Eq.~\eqref{tM}], we get two equations for the poles $\tilde{\omega}_\pm = \omega_\pm - i\gamma_\pm$, like in Appendix \ref{app:no PhC}, 
\begin{align}
    \e^{i\tilde{\omega}_\pm/\FSR} = \frac{1}{[r_\F(\tilde{\omega}_\pm) \pm t_\F(\tilde{\omega}_\pm)]r_\N}. \label{poles}
\end{align}
Using Eq.~\eqref{t r Fano}, we find $r_\F(\tilde{\omega}_-) - t_\F(\tilde{\omega}_-) = r_\D - t_\D$, which means that the $\omega^0_{n,-}$ family of resonances [Eq.~\eqref{w0m}], with linewidth $\gamma^0$, is unchanged. This is not the case for the ``$+$'' family since
\begin{align}
    r_\F(\tilde{\omega}_+) + t_\F(\tilde{\omega}_+) = (r_\D + t_\D)\frac{-\gF + i(\omF - \tilde{\omega}_+)}{\gF + i(\omF - \tilde{\omega}_+)}
\end{align}
and the pole equation becomes
\begin{align}
    \frac{-\gF + i(\omF - \tilde{\omega}_+)}{\gF + i(\omF - \tilde{\omega}_+)}\e^{i\tilde{\omega}_+/\FSR} = \e^{i\tilde{\omega}^0_{n, +}/\FSR}. \label{polesp}
\end{align}
For a small enough $\gF$, we expect to find a pole close to $\tilde{\omega}^0_{n, +}$ and another close to $\omF$ (with no imaginary part since we have assumed a lossless membrane). Writing $\tilde{\omega}_+$ as $\tilde{\omega}^0_{n, +} + \delta\tilde{\omega}$, with $\delta\tilde{\omega} = \delta\omega - i\delta\gamma$, Eq.~\eqref{polesp} becomes
\begin{align}\label{polesp2}
    2\gF  + i(\tilde{\Delta}^0_{n, +} + \delta\tilde{\omega}) = i(\tilde{\Delta}^0_{n, +} + \delta\tilde{\omega}) \e^{-i\delta\tilde{\omega}/\FSR},
\end{align}
with $\tilde{\Delta}^0_{n, +} = \tilde{\omega}^0_{n,+} -  (\omF - i\gF)$. 
In the case $\tilde{\Delta}^0_{n, +} \neq 0$,  we expand Eq.~\eqref{polesp2} at the first order in $\abs{\delta\omega}/\FSR, \abs{\delta\gamma}/\FSR$, take the real and imaginary part and solve the system of equations for $\delta\omega$ and $\delta\gamma$. We find
\begin{subequations}\allowdisplaybreaks\label{dw, dk}
\begin{align}
    \delta\omega&= \frac{ 2\gF \FSR ( \omega^0_{n,+} - \omF)}{( \omega^0_{n,+} - \omF)^2 + (\gamma^0  -\gF)^2} , \label{dw}\\
    \delta\gamma&= \frac{- 2\gF \FSR(\gamma^0  -\gF) }{( \omega^0_{n,+} - \omF)^2 + (\gamma^0  -\gF)^2},
\end{align}
\end{subequations}
such that the resonance is at frequency $\omega^0_{n, +} + \delta\omega$ and with linewidth $\gamma^0 + \delta \gamma$. One sees from Eq.~\eqref{dw, dk} that only the resonance $\omega^0_{n, +}$ closest to $\omF$ is shifted since $\abs{ \omega^0_{n,+} - \omF} \ge \FSR \gg \gF, \gamma^0$ for any other $n$.
Expanding around $\omF$, one finds that the pole is shifted to $\omF - \delta\omega + i\delta\gamma$, which requires $\delta\gamma$ to be negative, namely $\gF < \gamma^0$, which is consistent with the small $\gF$ limit considered here. 

In the case $\tilde{\Delta}^0_{n, +} =  0$ (namely $\omega_{n, +}^0 = \omF$ and $\gamma^0 = \gF$), Eq.~\eqref{polesp2} becomes
\begin{align}\label{polesp3}
    2\gF  + i \delta\tilde{\omega} = i\delta\tilde{\omega}\left(1 - i\frac{\delta\tilde{\omega}}{\FSR}\right),
\end{align}
where we now have to go to the second order in $\delta\tilde{\omega}/\FSR$, and we see that Eq.~\eqref{polesp3} only makes sense if $\gF/\FSR \ll 1$, which is consistent with our initial assumption. We find 
\begin{equation}\label{dw, dk 0}
    \delta\omega = \pm \sqrt{2\FSR\gF},\quad\delta\gamma = 0.
\end{equation}
This corresponds to an avoided crossing of the frequencies and a crossing of the loss rates. If we now compare to the normal mode analysis in the TM case (Sec.~\ref{sec:modelA:normal modes}) and in the RM case (Sec.~\ref{sec:modelB:normal modes}), we also expect an avoided crossing of two resonance frequencies, with a splitting $2\lambda_0$, giving the expression of $\lambda_0$ from Table~\ref{tab:param matching},
\begin{equation}
    \lambda_0 = \sqrt{2\FSR\gF}.\label{lambda0}
\end{equation}

\begin{figure}[tb]
    \includegraphics[width=\linewidth]{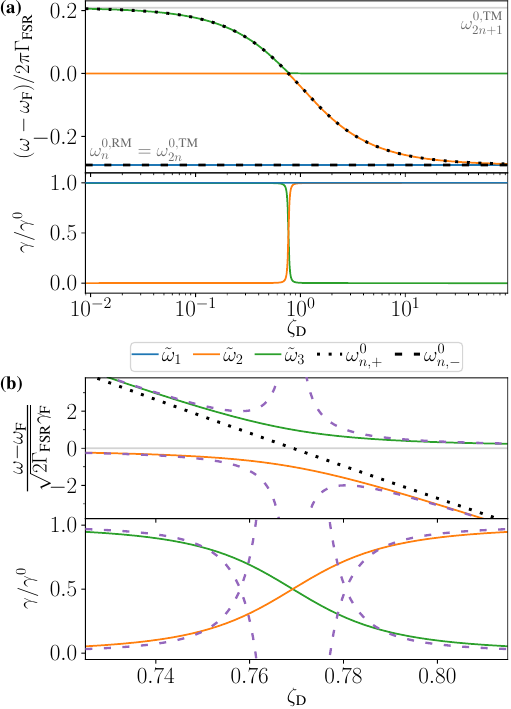}
    \caption{\label{fig:tm zD}
        Resonances $\tilde{\omega}_j = \omega_j - i\gamma_j$, with $j = 1, 2, 3$ of 
        the transfer-matrix model as functions of the polarizability $\zD$ of the unpatterned membrane. 
        Panel (b) is a zoom-in, in linear scale, of panel (a) in the area where $\tilde{\omega}_2$ and  $\tilde{\omega}_3$ are almost equal. The solid lines were obtained by numerically solving Eq.~\eqref{poles}. The thick dotted and dashed black lines are the resonances for an unpatterned membrane [Eq.~\eqref{w0pm}] and the gray horizontal lines indicate the FP resonances in the limits $\zD \to 0$ [Eq.~\eqref{wres_TM}] and $\zD \to \infty$ [Eq.~\eqref{wres_RM}]. In panel (b), the purple dashed lines correspond to the analytical approximations $\tilde{\omega}_{n, +}^0 + \delta\tilde{\omega}$, $\omF - \delta\tilde{\omega}$ [Eq.~\eqref{dw, dk}]. The parameters are $L = 50$ \si{\micro\meter}, $\zN = 50$,  and $\omF/2\pi = 192.75$ \si{\tera\hertz}, $\gF/\gamma^0 = 0.5$.
    }
\end{figure}

We have plotted in Fig.~\subfigref{fig:tm zD}{a} the three poles $\tilde{\omega}_j = \omega_j - i\gamma_j$ closest to $\omF$ as functions of the polarizability $\zD$. The dotted and dashed black lines indicate respectively $\omega^0_{n, +}$ and $\omega^0_{n, -}$, the resonance frequencies of the setup in the limit $\gF\to 0$ [Eq.~\eqref{w0pm}]. We see that $\tilde{\omega}_{n, +}^0$ and $\tilde{\omega}_{n, -}^0$ correspond to two distinct resonances of a FP cavity of length $2L$ at $\zD\to 0$ and ${\omega}_{n, +}^0$ progressively gets closer to ${\omega}_{n, -}^0$, to eventually reach the same frequency, corresponding to a FP resonance of a cavity of length $L$ [Eq.~\eqref{wres_RM}].
We have chosen a value of $\omF$ far away (compared to $\gF$) from the FP resonances in the fully transparent and fully reflective membrane limits. As a consequence, at both large and low polarizabilities, the resonance frequencies are far from each other, such that the interaction introduced by $\gF$ is negligible and each of the poles corresponds either to the bare photonic-crystal mode $\tilde{\omega}_\F$ or to one of the poles of the setup without photonic crystal, $\tilde{\omega}_{n, +}^0$ or $\tilde{\omega}_{n, -}^0$.  
This is no longer the case around the polarizability at which $\omega^0_{n, +}$ crosses $\omF$. 
While $\tilde{\omega}_1$ always perfectly coincides with $\tilde{\omega}^0_{n, -}$, there is an avoided crossing between frequencies $\omega_2$ and $\omega_3$ and a crossing of the loss rates $\gamma_2$ and $\gamma_3$. To make this more visible, Fig.~\subfigref{fig:tm zD}{b} shows a zoomed-in portion of Fig.~\subfigref{fig:tm zD}{a} around the crossing point of  $\omega^0_{n, +}$ and $\omF$. The dashed purple lines indicate the poles computed with the approximation from Eq.~\eqref{dw, dk}, which, as expected, stops being valid in the vicinity of the crossing, but is in good agreement with the exact values otherwise. Finally, we see that the distance between $\omega_2$ and $\omega_3$ at the avoided crossing is in excellent agreement with Eq.~\eqref{dw, dk 0}.

\begin{figure*}
    \includegraphics[width=\linewidth]{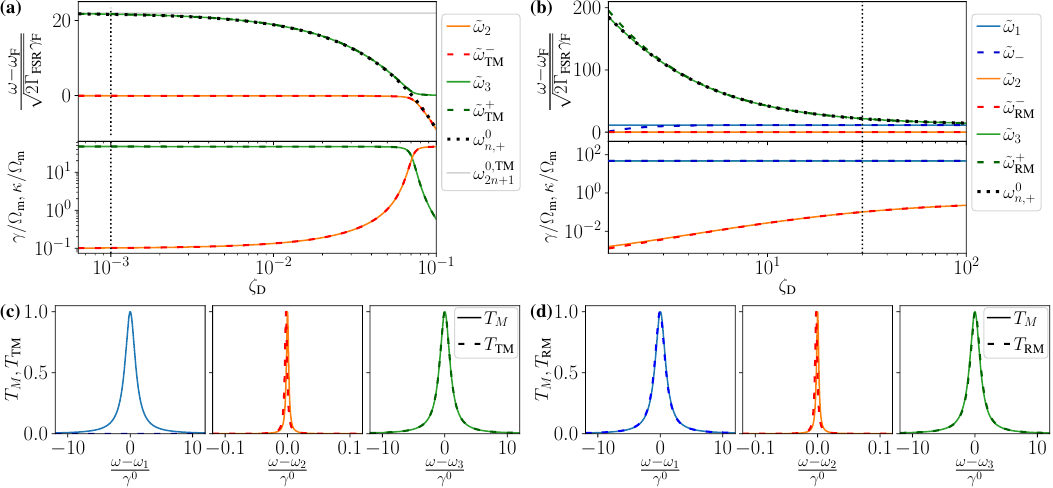}
    \caption{\label{fig:tm vs cm} Comparison of the transfer matrix and coupled-mode models: (a), (c) TM case and (b), (d) RM case. 
        (a), (b) Resonances of the transfer-matrix model (solid lines) and of the corresponding coupled-mode model (dashed lines) as functions of the polarizability $\zD$.
        (c), (d) Transmission in energy through the setup around the resonance frequencies $\omega_j$, $j = 1,2,3$ of the transfer-matrix model at the values of $\zD$ indicated by the vertical dotted black lines in (a) and (b), respectively.
        The parameters depending on $\zD$ were computed using the expressions in Table~\ref{tab:param matching}, while the other parameters are the ones given in Table~\ref{tab:all params} for the TM case in (a), (c) and the RM case in (b), (d).
    }
\end{figure*}

We now conduct a consistency check of the expressions found for $\lambda_0$ and $\chi$, Eqs.~\eqref{lambda0} and \eqref{chi}, by examining the TM and RM cases separately.

\subsubsection{Transparent membrane limit}\label{app:lambda:TM}

In the TM case, the transmission (in amplitude) without mechanics is given by 
\begin{equation}\label{t_TM2}
    t_\TM(\omega_\las) = \frac{ i\kappa_a\Dd }{\Omega^+_\TM\Omega^-_\TM},
\end{equation}
see Eqs.~\eqref{eq:transmission_modelA} and \eqref{eq:Omega_pm_modelA}, where we have taken $\kd = 0$ to match the transfer matrix assumptions.
First, $t_\TM (\omd) = 0$ [Eq.~\eqref{t_TM2}] but $t_\tm(\omF - \zD\gF) = 0$ [Eq.~\eqref{tM}], so the two expressions can only match in the limit $\zD \to 0$, as expected. 
Second, the poles of $t_\TM$ correspond to the normal modes, with $\omega^\pm_\TM = \Ree(\Omega^\pm_\TM) + \omega_\las$ and $\kappa^\pm_\TM = -\Imm(\Omega^\pm_\TM)$ [Eq.~\eqref{eq:Omega_pm_modelA}].

If $\omd$ and $\oma$ are relatively far apart, we expect only a small shift of $\omega^\pm_\TM$ compared to the initial frequencies since $\lambda_0 \ll \oma, \omd$. In the limit $\abs{\dD_\TM} \gg \lambda_0$, 
\begin{equation}
    \omega^\pm_\TM \simeq \bar{\Delta}_\TM + \omega_\las\pm \dD_\TM \pm \frac{\lambda_0^2\dD_\TM }{2\dD_\TM^2 + 2\delta\kappa^2}. 
\end{equation}
Therefore, we want to find $\lambda_0$ such that
\begin{equation}\label{dw_matching_TM}
    \frac{ 4\gF \FSR \dD_\TM}{ 4(\dD_\TM)^2 + (\ka  -\gF)^2}  \simeq \frac{2(\lambda_0)^2\dD_\TM }{4(\dD_\TM)^2 + \ka^2},
\end{equation}
where we have replaced $\omega^0_{n, +} - \omF$ by $\oma - \omd = 2\dD_\TM$ and $\gamma^0$ by $\ka$ in Eq.~\eqref{dw}, and $\delta \kappa = \ka/2$ since $\kd = 0$. The two expressions cannot exactly match but give equivalent frequency shifts in the limit $\abs{\dD_\TM} \gg \ka, \gF$,
which gives Eq.~\eqref{lambda0}, assuming $\lambda_0 > 0$.
In the case $\dD_\TM = 0$ (namely  $\oma = \omd$), we find a frequency shift $\pm \sqrt{(\lambda_0)^2 - \ka^2/4}$ [Eq.~\eqref{eq:Omega_pm_modelA}], which gives the same expression for $\lambda_0$ as
Eq.~\eqref{lambda0} when compared to Eq.~\eqref{dw, dk 0} in the limit $\ka \ll \lambda_0$. Since $\ka /\lambda_0 = \sqrt{\FSR/8\zN^4\gF} $, this is a reasonable regime for good mirrors, $\zN \gg 1$, even if $\gF/\FSR \ll 1$.

Because of the assumption we made to find this result, we see that the transfer matrix approach and the coupled mode model only match at the first order in $\gF/\FSR$. Having $\gF/\FSR \ll 1$ is a reasonable assumption to take since the coupled-mode model assumes that the membrane mode only couples to a single mode of the cavity. 
Furthermore, there is a fundamental mismatch between the transfer matrix and coupled-mode approaches: we would need to have both $\kd = 0$ to respect the absence of losses when the light goes through the membrane (no energy losses in the transfer-matrix model) and $\kd = \gF$, see Eq.~\eqref{dw_matching_TM}. However, as shown in Figs.~\subfigref{fig:tm vs cm}{a} and \subfigref{fig:tm vs cm}{c}, the two approaches are nevertheless in good agreement since we are in the regime $\gF/\FSR \ll 1$. 

Like in Fig.~\subfigref{fig:tm zD}{a}, we focus on the three poles $\tilde{\omega}_j = \omega_j - i\gamma_j$ closest to $\omF$ as functions of the polarizability $\zD$, but for small polarizabilities only, namely in the transparent membrane regime. Figure~\subfigref{fig:tm vs cm}{a} shows that $\tilde{\omega}_2$ and $\tilde{\omega}_3$ are in excellent agreement with the poles $\tilde{\omega}^\pm_\TM$ [Eq.~\eqref{eq:Omega_pm_modelA}] of the coupled-mode model in the TM case, with the parameters obtained from the expressions in Table~\ref{tab:param matching}.
The transmission rates of both models, $T_\tm(\omega) = \abs{t_\tm(\omega)}^2$ [Eq.~\eqref{tM}] and $T_\TM(\omega) = \abs{t_\TM(\omega)}^2$ [Eq.~\eqref{t_TM2}], are plotted in Fig.~\subfigref{fig:tm vs cm}{c} around the resonance frequencies $\omega_j$, $j = 1,2,3$ for $\zD = 0.001$, as indicated by the vertical dotted line in Fig.~\subfigref{fig:tm vs cm}{a}. We get three transmission peaks with the transfer-matrix model and only two, as expected, from the coupled-mode model since we have neglected the first resonance, as it is far enough away to not influence the two relevant optical modes much. These two peaks are also in excellent agreement with the transfer matrix results. The peaks around $\omega_1$ and $\omega_3$ are broad, with linewidths close to $\gamma^0 = \ka \simeq \kappa^+_\TM$, while the peak around $\omega_2$ is much narrower, with a linewidth $\kappa^-_\TM$, where $\kappa^-_\TM/\ka = 7\times 10^{-3}$. This is expected since the detuning between $\oma$ and $\omd$ is much larger than $\lambda_0$ (and $\gF$), such that resonances 1 and 3 are mostly FP-cavity-like, while resonance 2 is mostly membrane-like, but the small hybridization with the cavity mode gives it a finite linewidth, despite having $\kd = 0$. Apart from the finite value of $\kd$, the parameters used in Fig.~\subfigref{fig:cooling}{b} in the TM case are the same as in Fig.~\subfigref{fig:tm vs cm}{c}, justifying that they are experimentally realistic parameters.

\subsubsection{Reflective membrane limit}\label{app:lambda:RM}

In the RM case, the transmission (in amplitude) for the two-cavity-mode model without mechanics is given by
\begin{equation}
    t_\RM(\omega_\las) = \frac{i\ka(\lambda_0^2 - 2\chi\Dd)}{\Omega^{+}_\RM\Omega^{-}_\RM\tilde{\Delta}_-},
\end{equation}
see Eqs.~\eqref{eq:transmission_modelB} and \eqref{eq:Omega_pm_RM}.
In the limit $\abs{\dD_\RM} \gg \lambda_0$, the resonances are at frequencies $\omega_0 - \chi$ and
\begin{equation}\label{dwRM}
    \omega^\pm_\RM \simeq \bar{\Delta}_\RM + \omega_\las\pm \dD_\RM \pm \frac{\lambda_0^2\dD_\RM }{2\dD_\RM^2 + 2\delta\kappa^2}, 
\end{equation}
to second order in $\lambda_0/\abs{\dD_\RM}$.

First, we have $t_\RM(\omega_\las) = 0$ for $\omega_\las = \omd - \lambda_0^2/2\chi$, while, for the transfer matrix approach, $t_\tm(\omF - \zD\gF) = 0$ [Eq.~\eqref{tM}]. Therefore, we want $\omF - \zD\gF =  \omd - \lambda_0^2/2\chi$, and since $\omF = \omd$, this means $\lambda_0^2 =  2\chi \zD\gF$, which is consistent with Eqs.~\eqref{w0}, \eqref{chi} and \eqref{lambda0}.
Second, the expression of $\lambda_0$ is also consistent with the one we obtain by comparing the frequency shift from Eqs.~\eqref{dw, dk} and \eqref{dwRM}, as in Appendix~\ref{app:lambda:TM}, but with $\omega^0_{n, +} - \omF = \omega_0 + \chi - \omd$. It is also consistent with Eq.~\eqref{dw, dk 0} in the case $\omega_0 + \chi = \omd$, in the limit $\ka \ll \lambda_0$. 
Like for the TM case, the transfer matrix approach and the coupled mode model only match to first order in $\gF/\FSR$.

Figures \subfigref{fig:tm vs cm}{b} and \subfigref{fig:tm vs cm}{d} evidence that, like in the TM case [Figs.~\subfigref{fig:tm vs cm}{a} and \subfigref{fig:tm vs cm}{c}], the RM coupled-mode model reproduces very well the transfer matrix resonances, but in the high polarizability limit, $\abs{\zD} \gg 1$. In this limit, the coupled-mode model consists of three modes and, therefore, it reproduces the three resonances closest to $\omF$. In Fig.~\subfigref{fig:tm vs cm}{b}, we see that a mismatch appear between the two models when $\zD$ approaches 1, which is mostly due to the breakdown of the approximation $\pi/2 - \thD \simeq 1/\zD$ in the expression of $\chi$ [see Eq.~\eqref{chi}]. Even if it does not appear to be an issue for the polarizability $\zD = 30$ [vertical dotted line in Fig.~\subfigref{fig:tm vs cm}{b}] that we used to get the parameters for the RM case (see Table~\ref{tab:all params}), we used the exact expression to get a better match with the resonance peaks of the transmission. The transmission peaks in Fig.~\subfigref{fig:tm vs cm}{d} are very similar to those in Fig.~\subfigref{fig:tm vs cm}{c} for the TM case since we have chosen the parameters such that $\kappa^-_\TM = \kappa^-_\RM$, see Table~\ref{tab:all params}.

\subsection{Consequence for the optomechanical couplings}\label{app:om couplings}

We use the transfer-matrix model to infer some of the optomechanical couplings. To this end, we replace the propagation lengths in the left and right propagation matrices [Eq.~\eqref{M_prop}] by $L+x$ and $L-x$, respectively, where $x = \sqrt{2}\xzpf q$ denotes the small displacement of the membrane. Here, $\xzpf = \sqrt{\hbar/2 m\Om}$ is the zero-point fluctuation amplitude, with $m$ the effective mass of the mechanical resonator. In the TM limit, this does not affect the cavity resonance \cite{Jayich2008Sep}, while in the RM limit, from Eq.~\eqref{wres_RM}, we expect to have $g_{a_\L, 0} = -g_{a_\R, 0} = \omega_0\xzpf/L$ \cite{Aspelmeyer2014Dec}, as given in Table~\ref{tab:param matching}. In the absence of any other kind of optomechanical coupling, we find that $t_\tm(\omega)$ and $t_\RM(\omega)$ (and the normal modes defined in Sec.~\ref{sec:modelB:normal modes}) remain unchanged to first order in the mechanical displacement $q$, as expected \cite{Jayich2008Sep}. 

However, it was reported in Ref.~\cite{WWoe2023} that the properties of the membrane mode depend on the mechanical displacement due to the deformations of the photonic crystal. Specifically, we have
\begin{equation}\label{App_omF}
\omF(q) \simeq \omF(0) - \sqrt{2}g_{d, 0} q,
\end{equation}
with
$\omF(0) = \omd$, and 
\begin{equation}\label{App_gmF}
\gF(q) \simeq \gF(0) + \sqrt{2} g_{\gamma, 0} q. 
\end{equation}
The former gives rise to the dispersive optomechanical coupling  $-\hbar \sqrt{2}g_{d,0}\d^\dagger\d \q$ in the Hamiltonians [Eqs.~\eqref{eq:H_modelA} and \eqref{eq:H_modelB}]. The latter modulates the coupling $\lambda$ [Eq.~\eqref{lambda0}] between the membrane and cavity modes, such that
\begin{equation}
    \lambda(q) \simeq \lambda_0 +\sqrt{2} g_{\gamma, 0} \frac{\lambda_0}{2\gF(0)} q.
\end{equation}
This is the motivation behind the coupling term $\hbar\sqrt{2}g_{\lambda, 0}(\a^\dagger\d + \d^\dagger \a)\q$ in the Hamiltonians [Eqs.~\eqref{eq:H_modelA} and \eqref{eq:H_modelB}], with 
\begin{equation}\label{gl0}
    g_{\lambda, 0 } = g_{\gamma, 0} \frac{\lambda_0}{2\gF},
\end{equation} 
as given in Table~\ref{tab:param matching}. Note that $\abs{g_{\lambda, 0}} > \abs{g_{\gamma, 0}}$ since $\lambda_0/2\gF = \sqrt{\FSR/2\gF}$ and $\gF \ll \FSR$. The optomechanical coupling $g_{d, 0}$ introduced by the Fano mode in the membrane is crucial to get a sizable coupling between the narrowest optical normal mode and the mechanics; see Eq.~\eqref{g0-}.

\subsection{Conclusion on the transfer-matrix approach}

With this transfer-matrix model, we have been able to interpolate from the TM regime to the RM regime by varying the polarizability of the membrane. By studying the properties of the obtained optical transmission, we have determined the relations between the parameters of the transfer-matrix model and the optical parameters of the coupled-mode models, as summarized in Table~\ref{tab:param matching}. This allowed us to better ground our parameters in experimental reality since the transfer-matrix parameters, such as the length of the cavity and the polarizability of the membrane and mirrors, are easier to estimate for experimental devices than the couplings involved in the coupled-mode models.

\section{Validity of the linearization}\label{app:validity}

The linearized dynamics considered in this work, see in particular Appendixes \ref{app:TMa} and \ref{app:RMa}, is valid when there exists a well-defined solution of the nonlinear system of equations describing the semiclassical steady state, Eqs.~\eqref{eq:ss_modelA} and \eqref{eq:ss_modelB} for the TM and RM models, respectively. This can be verified by applying the Routh-Hurwitz criterion \cite{DeJesus1987Jun}. Following the procedure outlined in Ref.~\cite{DeJesus1987Jun}, we find, as in Ref.~\cite{Monsel2024Apr}, that the relevant stability criterion when driving on the red sideband of the sideband-resolved normal mode with increasing laser powers is $\det(\bm A_\TMRM) > 0$  [Eqs.~\eqref{eq:A_TM} and \eqref{eq:A_RM}] and when $\det(\bm A_\TMRM)$ becomes negative, so does $\nfin$. Therefore, the system is stable for the parameters of Fig.~\ref{fig:cooling}. 

\begin{figure}[tb]
    \includegraphics[width=\linewidth]{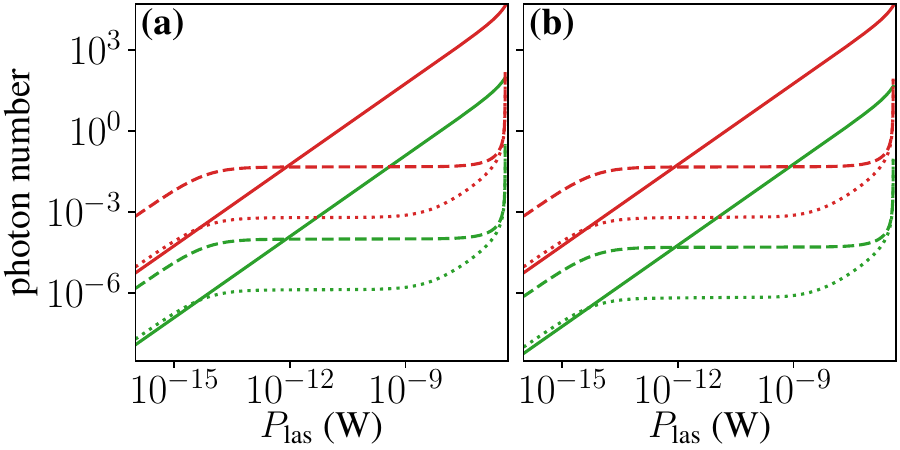}
    \caption{\label{fig:validity}
        Steady-state photon numbers as functions of the laser power, shown in green for the cavity modes and in red for the membrane mode, in (a) the TM case and (b) the RM case. The solid lines correspond to the classical steady state, $\abs{\bar{c}}^2$, while the fluctuations $\mean{\delta\hat{c}^\dagger\delta\hat{c}}$ are indicated by dashed lines for $T = 300$ K and dotted lines for $T = 4$ K, with $c = a,d$ in (a) and  $c = a_\L,a_\R, d$ in (b). In (b), the values for the left and right cavity modes are very similar such that the corresponding curves fully overlap. The parameters are the same as in Fig.~\subfigref{fig:cooling}{b}; see also Table~\ref{tab:all params}.
    }
\end{figure}

Furthermore, we neglect terms to second order in the fluctuations in the linearized Langevin equations \eqref{eq:linQLE_modelA} and \eqref{LinearQLE:LR}. To confirm the validity of this approximation, we compare in Fig.~\ref{fig:validity} the average photon numbers in the semiclassical steady state, $\abs{\bar{c}}^2$, to the average photon number in the fluctuations, $\mean{\delta\hat{c}^\dagger\delta\hat{c}}$, where $c = a,d$ in the TM case and $c = a_\L,a_\R, d$ in the RM case. The former is obtained by numerically solving Eqs.~\eqref{eq:ss_modelA} and \eqref{eq:ss_modelB}, while the latter is obtained by numerically solving the Lyapunov equations \eqref{eq:lyap_modelA} and \eqref{eq:Lyapunov_RM}.
Figure~\ref{fig:validity} shows that the average photon numbers in the semiclassical steady state are much larger than the photon numbers in the corresponding fluctuations for the laser powers relevant to achieve $\nfin < 1$ [see Fig.~\subfigref{fig:cooling}{b}], justifying the validity of the linearized regime. As expected, fluctuations are smaller at lower temperature and the membrane mode acquires a larger population than the cavity modes, since we are driving close to $\omega^-_\TMRM$, which is almost equal to $\omd$ for the considered parameters, see Table~\ref{tab:all params}.

\begin{table*}[t]
    \renewcommand{\arraystretch}{1.1}
    \setlength{\tabcolsep}{7pt}
    \begin{tabular}{llccc}
        \hline\hline
        Parameter             & Description                                                        & Unit               & TM case   & RM case  \\
        \hline
        \multicolumn{5}{l}{\textit{Mechanical mode}} \\
        
        $\Om/2\pi$            & Bare mechanical frequency                                          & \si{\mega\hertz}   & $2$       & $2$      \\
        $\gam/2\pi$           & Bare mechanical damping rate                                       & \si{\hertz}        & $0.02$    & $0.02$   \\
        $Q_\text{m}$          & Mechanical quality factor                                          & -                  & $10^8$    & $10^8$   \\
        $\xzpf$               & Zero-point fluctuation amplitude                                   & \si{\femto\meter}  & $1$       & $1$      \\
        
        \hline
        \multicolumn{5}{l}{\textit{Optics: transfer matrix parameters}}\\
        
        $L$                   & Distance between the membrane and the end mirrors                  & \si{\micro\meter}  & $50$      & $50$     \\
        $\zN = \tan(\thN)$    & Polarizability of the end mirrors                                  & -                  & $50$      & $50$     \\
        $\zD = \tan(\thD)$    & Polarizability of the unpatterned membrane                         & -                  & $0.001$   & $30$     \\
        $n_\TMRM$             & Number of the relevant FP resonance                                & -                  & $129$     & $64$     \\
        $\omF/2\pi$           & Fano resonance frequency                                           & \si{\tera\hertz}   & $193.31$  & $191.84$ \\
        $\gF/2\pi$            & Fano resonance linewidth                                           & \si{\mega\hertz}   & $10$      & $10$     \\
        $\FSR = c/2L$         & Free spectral range for a FP cavity of length $L$                  & \si{\tera\hertz}   & $3.00$    & $3.00$   \\
        
        \hline
        \multicolumn{5}{l}{\textit{Optical modes}}\\
        
        $\oma/2\pi$           & Cavity mode frequency in the TM case                               & \si{\tera\hertz}   & $193.37$  & -        \\
        $\omega_0/2\pi$       & Left and right cavity modes frequency in the RM case               & \si{\tera\hertz}   & -         & $191.89$ \\
        $\omd/2\pi$           & Membrane mode frequency                                            & \si{\tera\hertz}   & $193.31$  & $191.84$ \\
        $\dD_\TMRM/2\pi$      & Detuning between the relevant cavity resonance and the Fano mode   & \si{\giga\hertz}   & $33.4$    & $33.4$   \\
        $\lambda_0/2\pi$      & Coupling strength between the cavity and membrane modes            & \si{\giga\hertz}   & $3.09$    & $3.09$   \\
        $\chi/2\pi$           & Coupling strength between the left and right cavity modes          & \si{\giga\hertz}   & -         & $15.9$   \\
        $\ka/2\pi$            & Cavity mode loss rate                                              & \si{\mega\hertz}   & $95.4$    & $95.4$   \\
        $\kd/2\pi$            & Membrane mode loss rate                                            & \si{\mega\hertz}   & $0.5$     & $0.5$    \\
        
        \hline
        \multicolumn{5}{l}{\textit{Single-photon optomechanical couplings}}\\
        
        $g_{a,0}/2\pi$        & Cavity mode dispersive optomechanical coupling                     & \si{\kilo\hertz}   & $0$        & $3.84$   \\
        $g_{d,0}/2\pi$        & Membrane mode dispersive optomechanical coupling                   & \si{\kilo\hertz}   & $-0.182$   & $-0.182$ \\
        $g_{\gamma,0}/2\pi$   & Membrane mode dissipative optomechanical coupling                  & \si{\kilo\hertz}   & $0.321$    & $0.321$  \\
        $g_{\lambda,0}/2\pi$  & Mechanical modulation of the optical coupling $\lambda_0$          & \si{\kilo\hertz}   & $49.6$     & $49.6$   \\
        
        \hline
        \multicolumn{5}{l}{\textit{Normal mode with the narrowest linewidth}}\\
        
        $\omega^-_\TMRM/2\pi$ & Resonance frequency                                                & \si{\tera\hertz}   & $193.31$  & $191.84$ \\
        $\kappa^-_\TMRM/\Om$  & Effective sideband resolution                                      & -                  & $0.35$    & $0.35$   \\
        $\abs{g_{\TMRM,0}^-}/\Om$   & Effective single-photon ultrastrong-coupling ratio           & -                  & $0.0024$  & $0.0024$ \\
        
        \hline\hline
    \end{tabular}
    
    \caption{\label{tab:all params}
        Values of the parameters of all models for the TM and RM cases; see also Sec.~\ref{sec:exp} and Appendix~\ref{app:params}. These are the parameters used in Figs.~\subfigref{fig:cooling}{b}, \subfigref{fig:cooling}{c}, \subfigref{fig:tm vs cm}{c}, \subfigref{fig:tm vs cm}{d}, and \ref{fig:validity}.
    }
\end{table*}

\section{TM coupled-mode model} \label{app:TM}   

This appendix provides technical details on the calculation of the steady-state phonon number and the optical transmission in the TM case.

\subsection{Linearized dynamics and Lyapunov equation}\label{app:TMa}

The aim is to determine, using covariance-matrix calculations, the steady-state second moments of the linearized fluctuations, from which the final phonon occupation of the mechanical mode can be extracted.

By introducing optical quadratures $\dX_c=(\delta\hat c+\delta\hat c^\dagger)/\sqrt{2}$ and
$\dP_c=(\delta\hat c-\delta\hat c^\dagger)/(i\sqrt{2})$ for $c\in\{a,d\}$, Eq.~\eqref{eq:linQLE_modelA} can be rewritten as
$\dot{\bm Y}_{\rm TM}=\bm A_{\rm TM}\bm Y_{\rm TM} + \bm n_{\rm TM}$, with $\bm Y_{\rm TM}=(\dX_a,\dP_a,\dX_d,\dP_d,\dq,\deltp)^\mathsf{T}$ and
\allowdisplaybreaks
\begin{align}
    \bm A_{\rm TM}\!&=\!{\small\setlength{\arraycolsep}{1.5pt}
    \begin{pmatrix}
        -\ka     & \Delta_{a,0}         & 0          & \lambda    & -2\Imm(g_a) & 0\\
        -\Delta_{a,0}     & -\ka         & -\lambda   & 0          & 2\Ree(g_a) & 0\\
        0     & \lambda      & -\kd       & \Dd        & -2\Imm(g_d) & 0\\
        -\lambda & 0            & -\Dd       & -\kd       & 2\Ree(g_d) & 0\\
        0     & 0            & 0          & 0          & 0          & \Om\\
        2\Ree(g_a) & 2\Imm(g_a) & 2\Ree(g_d) & 2\Imm(g_d) & -\Om   & -\gam
    \end{pmatrix}},\nonumber\\
\bm n_{\rm TM} \!&=\!{\small
\begin{pmatrix}
    \sqrt{\kappa_a}\,(\hat X_{\rm in,L}+\hat X_{\rm in,R})\\
    \sqrt{\kappa_a}\,(\hat P_{\rm in,L}+\hat P_{\rm in,R})\\
    \sqrt{2\kappa_d}\,\hat X_{d,\rm in}\\
    \sqrt{2\kappa_d}\,\hat P_{d,\rm in}\\
    0\\
    \sqrt{\gam}\,\hat \xi
\end{pmatrix}},\label{eq:A_TM}
\end{align}

In the noise vector $\bm n_{\rm TM}$, the optical input quadratures are defined as
\begin{equation}
\hat X_{\rm in,\ell}
=
\frac{\hat a_{\rm in,\ell}+\hat a_{\rm in,\ell}^\dagger}{\sqrt{2}},
\qquad
\hat P_{\rm in,\ell}
=
\frac{\hat a_{\rm in,\ell}-\hat a_{\rm in,\ell}^\dagger}{i\sqrt{2}},
\end{equation}
with  $\ell \in\{\L, \R\}$, and
\begin{equation}\label{Xind, Pind}
\hat X_{d,\rm in}
=
\frac{\hat d_{\rm in}+\hat d_{\rm in}^\dagger}{\sqrt{2}},
\qquad
\hat P_{d,\rm in}
=
\frac{\hat d_{\rm in}-\hat d_{\rm in}^\dagger}{i\sqrt{2}}.
\end{equation}

The steady-state covariance matrix $\bm V_{\rm TM}$, whose elements are defined as 
\begin{equation}
\bm V_{{\rm TM},ij}
=
\frac{1}{2}
\left\langle
\bm Y_{{\rm TM},i} \bm Y_{{\rm TM},j}+\bm Y_{{\rm TM},j} \bm Y_{{\rm TM},i}
\right\rangle,
\label{Vij_def}
\end{equation}
satisfies the Lyapunov equation
\begin{equation}
\bm A_{\rm TM}\bm V_{\rm TM}+\bm V_{\rm TM}\bm A_{\rm TM}^\mathsf{T}+\bm B_{\rm TM}=0.
\label{eq:lyap_modelA}
\end{equation}
with the diffusion matrix defined through
\(
\bm B_{{\rm TM},ij}=\frac{1}{2}
\left\langle
\bm n_{{\rm TM},i}\bm n_{{\rm TM},j}
+
\bm n_{{\rm TM},j}\bm n_{{\rm TM},i}
\right\rangle.
\)

Assuming local Markovian baths for each mode, $\bm B_{\rm TM}$ is diagonal with optical vacuum noise and mechanical thermal noise:
\begin{equation}
\bm B_{\rm TM}=\mathrm{diag}
\left[
\kappa_a,\kappa_a,
\kappa_d,\kappa_d,
0,
\gam(2\nth+1)
\right].
\end{equation}
The final phonon occupancy is obtained from the mechanical variances as
\begin{align}
\bar n_\mathrm{fin}&=\frac{1}{2}\left(\mean*{\dq^{\,2}}+\mean*{\deltp^{\,2}}-1\right) \nonumber\\
&=\frac{1}{2}\left(\bm V_{{\rm TM},55}+\bm V_{{\rm TM},66}-1\right).
\label{eq:nfin_modelA}
\end{align}

\subsection{Optical transmission}\label{app:TMb} 

In the TM case, the transmission amplitude for left input and right output is $t_\TM(\omega_\las)=\frac{\mean{\bout[\R]}}{\mean{\bin[\L]}}$, with no input field from the right, where the input $\bin[\L]=\alas+\ainL$ is the laser drive and vacuum fluctuations. The output fields are given by the usual input-output relations \cite{Gardiner1985Jun}
\begin{equation}
\bout[\L]=\bin[\L]-\sqrt{\kappa_a}\a, \quad
\bout[\R]=\ainR-\sqrt{\kappa_a}\a.
\end{equation}
At the mean-field level, this yields
\begin{equation}
t_\TM(\omega_\las)=-\sqrt{\kappa_a}\,\frac{\ba}{\alas},
\quad
T_\TM(\omega_\las)=|t_\TM(\omega_\las)|^2.
\label{eq:transmission_modelA}
\end{equation}
In the absence of internal loss channels (formally $\kd\rightarrow 0$ in this reduced model), energy conservation implies $R_\TM+T_\TM=1$, where $R_\TM = \abs{\mean{\bout[\L]}/\mean{\bin[\L]}}^2$ is the reflectivity on the left side. Intrinsic membrane losses, namely $\kd > 0$, lead to $R_\TM+T_\TM<1$.

\begin{widetext}
\section{RM coupled-mode model}\label{app:RM}   

This appendix provides technical details about the linearization of the dynamics, the calculation of the steady-state phonon number, and the optical transmission in the RM case.
    
\subsection{Linearized dynamics and Lyapunov equation}\label{app:RMa}

Here we provide details for the quantum Langevin equations in the RM case and their linearization, as well as the corresponding Lyapunov equation. According to Eq.~\eqref{eq:H_modelB}, the quantum Langevin equations in the Markovian regime \cite{Gardiner1985Jun, Genes2008Mar, Monsel2024Apr} in this case can be obtained as 

\begin{subequations}\allowdisplaybreaks
    \label{eq:QLE_modelB:LR}
    \begin{align}
        \dot{\a}_\L &= -(i\Delta_0+\ka)\aL
        -i\chi\aR-i\frac{\lambda_0}{\sqrt{2}}\d+i\sqrt{2}g_{a,0}\q\,\aL -ig_{\lambda,0}\q\,\d +\sqrt{2\ka}\left(\alas+\ainL\right),\\
        \dot{\a}_\R &= -(i\Delta_0+\ka)\aR
        -i\chi\aL-i\frac{\lambda_0}{\sqrt{2}}\d -i\sqrt{2}g_{a,0}\q\,\aR -ig_{\lambda,0}\q\,\d +\sqrt{2\ka}\ainR,\\
        \dot{\d} &= -(i\Delta_{d,0}+\kd)\d
        -i\frac{\lambda_0}{\sqrt{2}}(\aL+\aR) +i\sqrt{2}g_{d,0}\q\,\d -ig_{\lambda,0}\q(\aL + \aR) +\sqrt{2\kd}\din,\\
        \dot{\q} &= \Om\p,\\
        \dot{\p} &= -\Om\q-\gam\p
        +\sqrt{2}g_{a,0}\left(\aL^\dagger\aL-\aR^\dagger\aR\right) -g_{\lambda,0}\left[(\aL^\dagger + \aR^\dagger)\d + \d^\dagger(\aL + \aR)\right]
        +\sqrt{2}g_{d,0}\d^\dagger\d+\sqrt{\gam}\,\hat\xi.
    \end{align}
\end{subequations}
By introducing the symmetric and antisymmetric modes \(\a_\pm = (\aL \pm \aR)/\sqrt{2}\), Eq.~\eqref{eq:QLE_modelB:LR} can be rewritten as Eqs.~\eqref{eq:QLE_modelB:+-}.
We then linearize the dynamics by writing
\[
\a_{\L(\R)}=\ba_{\L(\R)}+\da_{\L(\R)},\qquad
\d=\bd+\dd,\qquad
\q=\bq+\dq,\qquad
\p=\bar p+\deltp,
\]
where the steady-state mean values are obtained as
\begin{subequations}\allowdisplaybreaks\label{eq:ss_modelB}
    \begin{align}
        \baL  &=  \frac{\sqrt{2\ka}\left[(i\DaR + \ka)(i\Dd + \kd) + \lambda^2/2 \right] \alas}{(i\Delta_0 + \ka)^2(i\Dd + \kd) + \lambda^2(i\Delta_0 + \ka) + \chi^2(i\Dd + \kd) - i\chi\lambda^2} , \\
        \baR  &=  \frac{-\sqrt{2\ka}\left[i\chi(i\Dd + \kd) + \lambda^2/2 \right] \alas}{(i\Delta_0 + \ka)^2(i\Dd + \kd) + \lambda^2(i\Delta_0 + \ka) + \chi^2(i\Dd + \kd) - i\chi\lambda^2} , \\
        \bd  &=  \frac{-\sqrt{\ka}\lambda\left[i(i\DaR + \ka) +\chi \right] \alas}{(i\Delta_0 + \ka)^2(i\Dd + \kd) + \lambda^2(i\Delta_0 + \ka) + \chi^2(i\Dd + \kd) - i\chi\lambda^2} , \\        
        \bq &= \frac{\sqrt{2}}{\Om}\left\{g_{a,0} ( \abs{\baL}^2- \abs{\baR}^2)  + g_{d,0} \abs{\bd}^2-g_{\lambda,0}\left[\frac{\baL^* + \baR^*}{\sqrt{2}}\bd + \bd^*\frac{\baL + \baR}{\sqrt{2}}\right]\right\}, \quad \bar{p} = 0.
    \end{align}
\end{subequations}
Here $\DaL = \Delta_0 -\sqrt{2}g_{a,0}\bq$, $\DaR = \Delta_0 +\sqrt{2}g_{a,0}\bq$, $\Dd = \Delta_{d,0} - \sqrt{2}g_{d,0}\bq$, and $\lambda = \lambda_0 + \sqrt{2}g_{\lambda, 0}\bq$. Note that $(i\DaL + \ka)(i\DaR + \ka) \simeq (i\Delta_0 + \ka)^2$ is expanded up to the first order in $\bq$. The validity of this linearization is discussed in Appendix~\ref{app:validity}.

In this way, Eq.~\eqref{eq:QLE_modelB:LR} can be linearized as 
\begin{subequations}\allowdisplaybreaks\label{LinearQLE:LR}
    \begin{align}
        \delta\dot{\a}_\L &= -(i\DaL+\ka)\daL
        -i\chi\daR-i\frac{\lambda}{\sqrt{2}}\dd+i\sqrt{2}\gaL\dq +\sqrt{2\ka}\ainL,\\
        \delta\dot{\a}_\R &= -(i\DaR+\ka)\daR
        -i\chi\daL-i\frac{\lambda}{\sqrt{2}}\dd +i\sqrt{2}\gaR\dq +\sqrt{2\ka}\ainR,\\
        \delta\dot{\d} &= -(i\Dd+\kd)\dd
        -i\frac{\lambda}{\sqrt{2}}(\daL+\daR) +i\sqrt{2}g_d\dq +\sqrt{2\kd}\din,\\
        \delta\dot{\q} &= \Om\deltp,\\
        \delta\dot{\p} &= -\Om\dq-\gam\deltp+ \!\!\sum_{c=a_\L, a_\R, d}\!\!\sqrt{2}(g_c^*\delta\hat{c}+g_c\delta\hat{c}^\dagger) +\sqrt{\gam}\,\hat\xi,
    \end{align}
\end{subequations}
where $\gaL = g_{a,0}\baL -g_{\lambda,0}/\sqrt{2}\bd$,  $\gaR = -g_{a,0}\baR -g_{\lambda,0}/\sqrt{2}\bd$, $g_d = g_{d, 0}\bd - g_{\lambda,0}(\baL + \baR)/\sqrt{2}$. 

Similarly, for the symmetric and antisymmetric modes, we have $\ba_\pm = (\baL \pm \baR)/\sqrt{2}$ and $\da_\pm = (\daL \pm \daR)/\sqrt{2}$. So, we rewrite Eq.~\eqref{LinearQLE:LR} as
\begin{subequations}\allowdisplaybreaks\label{LinearQLE:pm}
    \begin{align}
        \delta\dot{\a}_- &= -(i\Delta_- +\ka)\da_- + ig_{+-}\da_+ +i\sqrt{2}g_{a_-}\dq+\sqrt{2\ka}\ain[-]
        ,\\
        \delta\dot{\a}_+ &= -(i\Delta_+ +\ka)\da_+ +ig_{+-}\da_--i\lambda\dd +i\sqrt{2}g_{a_+}\dq+\sqrt{2\ka}\ain[+],\\
        \delta\dot{\d} &= -(i\Dd+\kd)\dd
        -i\lambda\da_+ +i\sqrt{2}g_d\dq +\sqrt{2\kd}\din,\\
        \delta\dot{\q} &= \Om\deltp,\\
        \delta\dot{\p} &= -\Om\dq-\gam\deltp+ \!\!\sum_{c=a_-, a_+, d}\!\!\sqrt{2}(g_c^*\delta\hat{c}+g_c\delta\hat{c}^\dagger) +\sqrt{\gam}\,\hat\xi,
    \end{align}
\end{subequations}
where $\Delta_\pm = \Delta_0 \pm \chi$, $g_{+-} = \sqrt{2}g_{a, 0}\bq$, $g_{a_-} =g_{a,0}\ba_+ $, and $g_{a_+}  = g_{a,0}\ba_- -g_{\lambda,0}\bd$. The effective coupling $g_{+-}$ between $\da_+$ and $\da_-$ is expected to be much weaker than the other couplings, since it is not enhanced by the drive laser (unlike the effective optomechanical couplings). 

We now introduce the optical quadratures
\begin{equation}\label{AppB:quadrature}
\dX_c=\frac{\delta\hat c+\delta\hat c^\dagger}{\sqrt{2}},
\qquad
\dP_c=\frac{\delta\hat c-\delta\hat c^\dagger}{i\sqrt{2}},
\qquad
c\in\{a_-,a_+,d\}.
\end{equation}
For compactness, we denote
\(
\dX_{\pm} \equiv \dX_{a_{\pm}}\) and
\(\dP_{\pm} \equiv \dP_{a_{\pm}}
\).
In this quadrature basis, Eq.~\eqref{LinearQLE:pm} becomes
\begin{subequations}\allowdisplaybreaks
\label{eq}
\begin{align}
\delta\dot{\hat X}_-
&=
-\ka\dX_- +\Delta_-\dP_-
-g_{+-}\dP_+-2\Imm(g_{a_-})\dq
+\sqrt{2\ka}\hat X_{\rm in,-},
\\
\delta\dot{\hat P}_-
&=
-\Delta_-\dX_- -\ka\dP_-
+g_{+-}\dX_+
+2\Ree(g_{a_-})\dq
+\sqrt{2\ka}\hat P_{\rm in,-},
\\
\delta\dot{\hat X}_+
&=
-\ka\dX_+ +\Delta_+\dP_+
-g_{+-}\dP_-
+\lambda\dP_d-2\Imm(g_{a_+})\dq
+\sqrt{2\ka}\hat X_{\rm in,+},
\\
\delta\dot{\hat P}_+
&=
-\Delta_+\dX_+ -\ka\dP_+
+g_{+-}\dX_-
-\lambda\dX_d
+2\Ree(g_{a_+})\dq
+\sqrt{2\ka}\hat P_{\rm in,+},
\\
\delta\dot{\hat X}_d
&=
-\kd\dX_d+\Dd\dP_d
+\lambda\dP_+-2\Imm(g_{d})\dq
+\sqrt{2\kd}\hat X_{d,\rm in},
\\
\delta\dot{\hat P}_d
&=
-\Dd\dX_d-\kd\dP_d
-\lambda\dX_+
+2\Ree(g_d)\dq
+\sqrt{2\kd}\hat P_{d,\rm in},
\\
\delta\dot{\q}
&=
\Om\deltp,
\\
\delta\dot{\p}
&=
\sum_{c=a_-, a_+, d}\left[2\Ree(g_{c})\dX_c + 2\Imm(g_{c})\dP_c\right]
-\Om\dq-\gam\deltp
+\sqrt{\gam}\hat\xi.
\end{align}
\end{subequations}
Here, the input noise quadratures are defined as
\begin{equation}
\hat X_{{\rm in},\pm}
=
\frac{\ain[\pm]+\ain[\pm]^\dagger}{\sqrt{2}},
\qquad
\hat P_{{\rm in},\pm}
=
\frac{\ain[\pm]-\ain[\pm]^\dagger}{i\sqrt{2}},
\qquad
\hat X_{d,{\rm in}}
=
\frac{\din+\din^\dagger}{\sqrt{2}},
\qquad
\hat P_{d,{\rm in}}
=
\frac{\din-\din^\dagger}{i\sqrt{2}}.
\end{equation}

Equivalently, by collecting the fluctuation quadratures into
\begin{equation}
\bm Y_\RM
=
\left(
\dX_-,
\dP_-,
\dX_+,
\dP_+,
\dX_d,
\dP_d,
\dq,
\deltp
\right)^\mathsf{T},
\label{eq:Y_RM}
\end{equation}
the linearized Langevin equations can be written in the compact form
\begin{equation}
\dot{\bm Y}_\RM
=
\bm A_\RM\bm Y_\RM
+
\bm n_\RM,
\label{eq:compact_RM}
\end{equation}
where the drift matrix is
\begin{equation}
\bm A_\RM
=
\begin{pmatrix}
-\ka           & \Delta_-       & 0              & -g_{+-}   & 0              & 0          & -2\Imm(g_{a_-}) & 0    \\
-\Delta_-      & -\ka           & g_{+-}         & 0              & 0              & 0          & 2\Ree(g_{a_-})  & 0    \\
0              & -g_{+-}   & -\ka           & \Delta_+       & 0              & \lambda    & -2\Imm(g_{a_+}) & 0    \\
g_{+-}         & 0              & -\Delta_+      & -\ka           & -\lambda       & 0          & 2\Ree(g_{a_+})  & 0    \\
0              & 0              & 0              & \lambda        & -\kd           & \Dd        & -2\Imm(g_d)     & 0    \\
0              & 0              & -\lambda       & 0              & -\Dd           & -\kd       & 2\Ree(g_d)      & 0    \\
0              & 0              & 0              & 0              & 0              & 0          & 0               & \Om  \\
2\Ree(g_{a_-}) & 2\Imm(g_{a_-}) & 2\Ree(g_{a_+}) & 2\Imm(g_{a_+}) & 2\Ree(g_d)     & 2\Imm(g_d) & -\Om            & -\gam
\end{pmatrix}.
\label{eq:A_RM}
\end{equation}
The corresponding noise vector is
\begin{equation}
\bm n_\RM
=
\left(
\sqrt{2\ka}\,\hat X_{{\rm in},-},
\sqrt{2\ka}\,\hat P_{{\rm in},-},
\sqrt{2\ka}\,\hat X_{{\rm in},+},
\sqrt{2\ka}\,\hat P_{{\rm in},+},
\sqrt{2\kd}\,\hat X_{d,{\rm in}},
\sqrt{2\kd}\,\hat P_{d,{\rm in}},
0,
\sqrt{\gam}\,\hat\xi
\right)^\mathsf{T}.
\label{eq:n_RM}
\end{equation}

For optical vacuum noise and a mechanical thermal bath with occupation \(\nth\), the symmetrized noise correlations define the diffusion matrix \(\bm B_\RM\) as
\begin{equation}
\bm B_\RM
=
\mathrm{diag}
\left[
\ka,\ka,
\ka,\ka,
\kd,\kd,
0,
\gam(2\nth+1)
\right].
\label{eq:D_RM}
\end{equation}
The steady-state covariance matrix $\bm V_\RM$, as defined in Eq.~\eqref{Vij_def} but with $\bm Y_\RM$, obeys
\begin{equation}
\bm A_\RM\bm V_\RM
+
\bm V_\RM\bm A_\RM^{\mathsf T}
+
\bm B_\RM
=
0.
\label{eq:Lyapunov_RM}
\end{equation}
Solving Eq.~\eqref{eq:Lyapunov_RM} gives all steady-state second moments, as in the TM case. Since \(\dq\) and \(\deltp\) are the seventh and eighth components of \(\bm Y_\RM\), respectively, the final phonon occupancy is
\begin{equation}
\bar n_{\rm fin}
=
\frac{1}{2}
\left(
\mean*{\dq^{\,2}}+\mean*{\deltp^{\,2}}-1
\right)
=
\frac{1}{2}
\left(
\bm V_{\RM,77}+\bm V_{\RM,88}-1
\right).
\label{eq:nfin_RM}
\end{equation}

\subsection{Optical transmission}\label{app:RMb}

As in the TM case, the output field is given by the usual input-output relation \cite{Gardiner1985Jun} $\bout[\R]=\bin[\R]-\sqrt{2\ka}\aR$ and the mean-field transmission is calculated from
\begin{equation}
        t_\RM(\omega_\las)=\frac{\mean*{\bout[\R]}}{\alas}
        =-\sqrt{2\ka}\,\frac{\baR}{\alas},\qquad
        T_\RM(\omega_\las)=|t_\RM(\omega_\las)|^2.
        \label{eq:transmission_modelB}
\end{equation}
The spectrum typically shows three features associated with the hybrid optical resonances, whose separation and linewidths provide a direct diagnostic of the underlying coupling parameters.

\end{widetext}

\section{Parameters and experimental feasibility}\label{app:params}

All the parameters for the ground-state cooling study from Fig.~\ref{fig:cooling} are given in Table~\ref{tab:all params}. 
As discussed in Sec.~\ref{sec:exp}, to ensure that these parameters are realistic, especially $\lambda_0$ and $\chi$, we determined them based on the transfer-matrix model presented in Appendix~\ref{app:tm}, see also Table~\ref{tab:param matching}. 
Note that even if the rounded values of $\omega^-_\TMRM/2\pi$ are equal to $\omd/2\pi$ in the TM and RM cases, respectively, they actually differ by around  $0.1$ \si{\giga\hertz}, since $\Delta^-_\TMRM \simeq \Delta_{d,0} -\lambda_0^2/2\dD_\TMRM$ in the limit $\dD_\TMRM \gg \lambda_0$ relevant here, see Eqs.~\eqref{eq:Omega_pm_modelA} and \eqref{eq:Omega_pm_RM}.

For the optomechanical couplings of the membrane mode,  $g_{d,0}$ and $g_{\gamma, 0}$, we took the values of the dispersive and dissipative couplings of the Fano mode in the experimental device from Ref.~\cite{WWoe2023}, reducing them by a factor $10^4$ to stay in the linear regime. We then computed $g_{\lambda, 0}$ using Eq.~\eqref{gl0}.

\section{Trade-off between sideband resolution and efficiency}\label{app:efficiency}

\begin{figure}[b]
    \includegraphics[width=\linewidth]{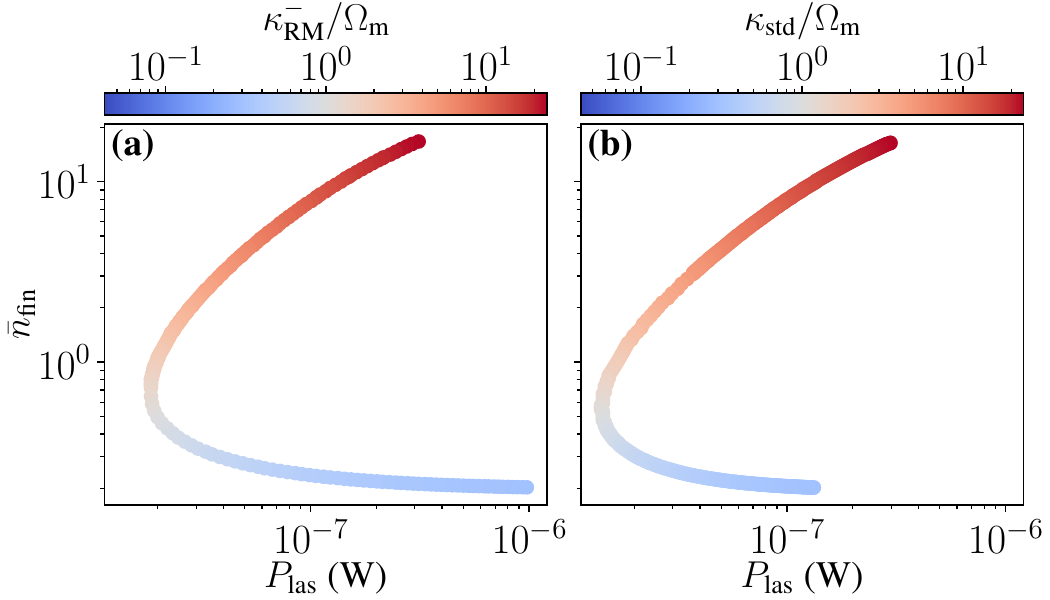}
    \caption{\label{fig:lasso} 
        (a) Optimized $\nfin$ as a function of the laser power $P_\las$ required to reach it, obtained by varying $\dD_\RM$, namely this is a parametric representation of the two curves in the bottom panel of Fig.~\subfigref{fig:cooling2}{b}. (b) Optimized $\nfin$ as a function of $P_\las$ for the equivalent standard optomechanical device: cavity with frequency $\omega^-_\RM$, with a single-photon optomechanical coupling given by $\abs{g_{\RM,0}^-}$ [Eq.~\eqref{g0-}] and a linewidth $\kappa_\text{std} = \kappa^-_\RM$. The colormap indicates the sideband resolution ratio.  
    }
\end{figure}

The trade-off between the minimum reachable $\nfin$ and the corresponding laser power, discussed in Sec.~\ref{sec:cooling} and in particular Fig.~\subfigref{fig:cooling2}{b}, is particularly visible when plotting the optimized $\nfin$ as a function of $P_\las$; see Fig.~\subfigref{fig:lasso}{a}. 
To gain a deeper understanding of this trade-off, we also plotted the same curve for an equivalent standard optomechanical device, that is a FP cavity with one movable end mirror, in  Fig.~\subfigref{fig:lasso}{b}. For a fair comparison, we used $\omega^-_\RM$ as the cavity bare frequency, and $\kappa^-_\RM$ as the optical loss rate to have the same sideband resolution ratio. Furthermore, we used $\abs{g_{\RM,0}^-}$ [Eq.~\eqref{g0-}] as the single-photon optomechanical coupling strength. The two setups reach the same values of $\nfin$, and both require the least laser power at the beginning of the sideband-resolved regime, for optical loss rates slightly smaller than the mechanical frequency. However, when the setup enters a regime of deeper sideband resolution, the required laser power increases more slowly for the standard setup. As a consequence, we can assert that the sharp increase in laser power is caused not only by the decreased optomechanical coupling [see Eq.~\eqref{g0-}, noting that $g_{d,0}$ is typically negative \cite{WWoe2023}], but also by a less efficient driving of the narrow normal mode.

\bibliography{cite.bib}
\end{document}